\documentstyle[12pt]{article}

\newcommand{\bmat}{\left(\begin{array}}
\newcommand{\emat}{\end{array}\right)}
\def\NPB#1#2#3{Nucl. Phys. B{#1} (19#2) #3}
\def\PLB#1#2#3{Phys. Lett. B{#1} (19#2) #3}

\def\PRD#1#2#3{Phys. Rev. D{#1} (19#2) #3}
\def\PRL#1#2#3{Phys. Rev. Lett. {#1} (19#2) #3}

\def\Deq#1{\mbox{$D$=#1}}
\def\Neq#1{\mbox{$N$=#1}}

\def\pim{{\rm Im\,}}

\def\yzero{\smash{\hbox{$y\kern-4pt\raise1pt\hbox{${}^\circ$}$}}}

\def\a{\alpha}
\def\b{\beta}
\def\g{\gamma}
\def\d{\delta}
\def\Om{\Omega}
\def\om{\omega}
\def\th{\theta}
\def\vt{\vartheta}
\def\vphi{\varphi}
\def\-{\hphantom{-}}
\def\ov{\overline}
\def\s2{\frac{1}{\sqrt2}}

\def\oh{\frac{1}{2}}
\def\beq{\begin{equation}}
\def\eeq{\end{equation}}
\def\beqa{\begin{eqnarray}}
\def\eeqa{\end{eqnarray}}

\def\Tr{{\rm Tr \,}}
\def\diag{{\rm diag \,}}
\def\IF{\relax{\rm I\kern-.18em F}}
\def\II{\relax{\rm I\kern-.18em I}}
\def\IP{\relax{\rm I\kern-.18em P}}
\def\IC{\relax\hbox{\kern.25em$\inbar\kern-.3em{\rm C}$}}
\def\IR{\relax{\rm I\kern-.18em R}}

\def\vac{|0 \rangle}

\def\cc{{\cal C}}
\def\ck{{\cal K}}

\def\cg{{\cal G}}

\def\cam{{\cal M}}

\def\cz{{\cal Z}}

\def\Dsl{\,\raise.15ex\hbox{/}\mkern-13.5mu D} 
\def\IZ{Z\kern-.4em  Z}
\def\id{{\rm I}}

\topmargin
-1.5cm
\textwidth
15.5cm
\textheight
23.5cm
\oddsidemargin
0.7cm
\evensidemargin
1.2cm

\begin{document}

\makeatletter
\@addtoreset{equation}{section}
\makeatother
\renewcommand{\theequation}{\thesection.\arabic{equation}}
\pagestyle{empty}
\rightline{FTUAM-98/4; IFT-UAM/CSIC-98-4; UCVFC-DF/12-98}
\rightline{\tt hep-th/9804026}
\vspace{0.5cm}
\begin{center}
\LARGE{
D=4, N=1, Type IIB Orientifolds \\[10mm]}
\large{G.~Aldazabal$^1$,
A. Font$^2$,
L.~E.~Ib\'a\~nez$^3$
and G. Violero$^3$\\[2mm]}
\small{
$^1$ CNEA, Centro At\'omico Bariloche,\\[-0.3em]
8400 S.C. de Bariloche, and CONICET, Argentina.\\[1mm]
$^2$ Departamento de F\'{\i}sica, Facultad de Ciencias,
Universidad Central de Venezuela \\[-0.3em]
A.P. 20513, Caracas 1020-A, Venezuela. \\[-0.3em]
and \\[-0.3em]
Centro de Astrof\'{\i}sica Te\'orica, Facultad de Ciencias,\\[-0.3em]
Universidad de Los Andes, Venezuela.\\[1mm]
$^3$ Departamento de F\'{\i}sica Te\'orica C-XI
and Instituto de F\'{\i}sica Te\'orica  C-XVI,\\[-0.3em]
Universidad Aut\'onoma de Madrid,
Cantoblanco, 28049 Madrid, Spain.
\\[9mm]}
\small{\bf Abstract} \\[7mm]
\end{center}

\begin{center}
\begin{minipage}[h]{14.0cm}
We study different aspects of the construction of $D$=4, $N$=1 type IIB
orientifolds based on toroidal $Z_N$ and $Z_M\times Z_N$, $D$=4 orbifolds.
We find that tadpole cancellation conditions are in general
more constraining than in six dimensions and that the
standard Gimon-Polchinski orientifold projection leads to the
impossibility of  tadpole cancellations in a number of
$Z_N$ orientifolds with even $N$ including $Z_4$, $Z_8$,
$Z_8'$ and $Z_{12}'$.
We construct  $D$=4, $Z_N$ and
$Z_N\times Z_M$  orientifolds  with  different configurations of
9-branes, 5-branes and 7-branes, most of them chiral.
Models including the analogue of discrete
torsion are constructed and shown to have features
previously conjectured on the basis of F-theory
compactified on four-folds.
Different properties of the
$D$=4, $N$=1 models obtained are discussed including  their
possible heterotic duals and effective low-energy
action. These models have in general more than one anomalous $U(1)$
and the anomalies are cancelled by a $D$=4 generalized
Green-Schwarz mechanism involving dilaton and moduli fields.

\end{minipage}
\end{center}
\newpage
\setcounter{page}{1}
\pagestyle{plain}
\renewcommand{\thefootnote}{\arabic{footnote}}
\setcounter{footnote}{0}

\section{Introduction}

Although all \Deq10, \Neq1 superstring theories are thought to be
equally consistent,
only the space of classical \Deq4, \Neq1
vacua of the  $E_8\times E_8$ heterotic string has been
studied in some detail. On the contrary, perturbative \Deq4, \Neq1
vacua of type I  have remained very much unexplored because
of different reasons. Compactification of type I theory
on a Calabi-Yau threefold with standard embedding in the
gauge degrees of freedom
gives rise to consistent (order by order in $\alpha^\prime$)
\Deq4, \Neq1 classical vacua but the gauge group
$SO(26)\times U(1)$ is non-chiral. The construction
of fully-fledged four-dimensional type I strings  is relatively recent.
Type I strings can be understood as an orbifold
(orientifold) of type IIB closed strings with respect to the
world-sheet parity operation $\Omega$ \cite{sagnotti,hor}.
Type IIB \Neq1, \Deq6  orientifolds  have been constructed
in the last few years in refs.~\cite{bs,gp,dp1,gj,dp2,dp3,bz,blum,kak4}.
A crucial ingredient in this
construction is the existence of D$p$-branes \cite{poldb} on which
open type I strings must end. Their presence from
this perspective is enforced upon us by the tadpole
cancellation conditions which guarantee the cancellation of
gauge and gravitational anomalies.
Type I vacua are not only worth constructing by themselves but also
because they are supposed to be S-dual to strongly coupled $SO(32)$
heterotic vacua and hence there is a hope to get information about
non-perturbative heterotic physics.

Our knowledge of the structure of \Deq4, \Neq1 type IIB
orientifolds is much less complete, although
some examples have been
constructed \cite{bl,ang,kak1,kak2,kak3,fin}
and general conditions for
tadpole cancellation in $Z_N\times Z_M$ type IIB orientifolds
have been recently presented
\cite{zwart,odri}. In this paper we
undertake a systematic study of \Deq4, \Neq1 type IIB orientifolds
and extend previous work in several different directions.
We present a detailed study of tadpole cancellation conditions
for general \Deq4, $Z_N$ orientifolds and find the surprising result
that the usual Gimon-Polchinski orientifold projection \cite{gp}
leads to the impossibility of tadpole cancellation for
most of the even order $Z_N$, \Deq4 orientifolds. In particular
this is the case for the $Z_4$, $Z_8$, $Z_8'$ and $Z_{12}'$
orientifolds. This is to be contrasted with the \Deq6 case in which
all $Z_N$ actions have a Chan-Paton
realization compatible with tadpole cancellation \cite{gj,dp2}.

We explicitly construct the massless spectrum of all
\Deq4 consistent orientifolds with at most one set of
5-branes sitting at the fixed point at the origin.
 We find particularly useful a Cartan-Weyl
realization of the unitary matrices $\gamma _{\theta,p}$
which induce the orbifold action on the D$p$-brane degrees
of freedom. In this formulation the massless spectra is
found in a straightforward manner reminiscent of the
computation of massless spectra in heterotic orbifolds.
Another feature shown (which also occurs in \Deq6) is
that tadpole cancellation conditions vary depending on
what particular fixed points do host 5-branes. We also discuss
the effect of the addition of (quantized)
Wilson lines on the 9-brane sector
and the T-dual of this which is the distribution of 5-branes
on deferent fixed points.
T-duality also maps continuous Wilson lines to the emission
of sets of 5-branes from the fixed points to the bulk of the
orbifold.

In refs.~\cite{polten,dp3,bz,blum} it was shown that in even order \Deq6
orientifolds there are alternative ways to project
the closed string sector with respect to the world-sheet parity
$\Omega $. This is an equivalent of the discrete torsion
degree of freedom already found in heterotic orbifolds \cite{vafadt, fiq}.
We  construct  the first \Deq4, \Neq1
orientifolds with these characteristics. This class of models
is interesting since their existence was  conjectured on the
basis of F-theory compactified on four-folds \cite{gopm}.

We also examine the construction of candidate heterotic duals
for type IIB \Neq1 orientifolds. The Cartan-Weyl basis for
the gauge embedding of the twists on D-branes mentioned above
is specially useful in identifying the candidate heterotic
duals. We argue that the heterotic duals of the class of
type IIB orientifolds discussed in this paper are in general
non-perturbative heterotic orbifolds. These are orbifolds
of the class introduced in ref.~\cite{afiuv} in which the embedding of the
twist in the $Spin(32)/Z_2$ lattice violates the standard modular invariance
constraints of perturbative orbifolds. The gauge interactions
and charged chiral fields from the type I (99) sector are
mapped into the untwisted sector of the heterotic orbifold.
The (55) and (95) sectors map into  non-perturbative degrees of freedom
associated to small instanton effects. We describe
a few examples of candidate heterotic duals.

We finally study some aspects of the effective low-energy
Lagrangian of IIB orientifolds, focusing on those
with both 9-branes and one sector of 5-branes.
A truncation to four dimensions of the \Deq10 Lagrangian
and symmetry arguments allow us to obtain some generic qualitative
information on the form of the K\"ahler potential, superpotential
and gauge kinetic functions.
Among the generic features is the presence of several
anomalous $U(1)$'s whose anomalies are cancelled by a
generalized Green-Schwarz mechanism in which both
dilaton and moduli fields are involved. This is to be contrasted
to perturbative heterotic vacua which have at most one
anomalous $U(1)$.

\section{$D=4$, $N=1$,  Type IIB  Orientifolds}
\label{basics}

In this chapter we summarize the basic ingredients \cite {sagnotti,bs,gp} 
and notation needed in the construction of orientifolds and generalize them to the
\Deq4 case. 

An orientifold is a generalization of an orbifold in which a
toroidally compactified theory is divided out by an internal
discrete symmetry $G_1$ such as $Z_N$.  In type IIB string theory there is
symmetry operation $\Omega$ that exchanges left and right world sheet movers.
Gauging away this symmetry produces the orientifold and leads to the emergence
of non-oriented surfaces spanned by  string propagation.
Generically, the $\Omega$ parity transformation
can be accompanied by  other internal or space time symmetry operations.
Examples of this are considered in section 4.
The  complete orientifold group can thus be written as
$G_1+ {\Omega} G_2$ with ${\Omega}h {\Omega} h' \in G_1$ for $h,h' \in
G_2$ \cite{gp}.

In most of this article we will be mainly
concerned  with $G_1=G_2$ and $G_1=Z_N$
or $G_1=Z_N\times Z_M$ actions on $T^6$ in type IIB string theory
(in section 4 we will consider cases with $G_2\not=G_1$).
The $Z_N$ orbifold action is realized by powers of the twist generator
$\theta $ ($\theta^N=1$) which can be written in the form
\begin{equation}
\theta= \exp (2i\pi (v_1J_{45}+v_2J_{67}+v_3J_{89}))
\label{twist}
\end{equation}
where $J_{mn}$ are $SO(6)$ Cartan generators.
In terms of the complex bosonic coordinates $Y_1 = X_4+iX_5$, $Y_2 = X_6+iX_7$
 and $Y_3 = X_8+iX_9$  that parametrize the torus, $\th$ acts diagonally as
\begin{equation}
\theta^k Y_i  = {\rm e}^{2i\pi kv_i}Y_i
\label{thdi}
\end{equation}
Similarly, we define complex fermionic fields $\psi^i$ as
$\psi^1= \psi^4+i\psi^5$, etc..
It is convenient to define a twist vector
$v= (v_1, v_2, v_3)$ associated to $\th$. \Neq1 supersymmetry requires
$\pm v_1 \pm v_2 \pm v_3= 0$ for some choice of signs \cite{dhvw}.
$Z_N \times Z_M$ actions are described in a similar way \cite{fiq}.
In this case we have $\theta$ and $\omega$ generators
whose associated twist vectors are
$v_\th= {\frac{1}N }(1,-1,0)$ and $v_\om={ \frac{1}M}(0, 1,-1)$.
In what follows we focus on $Z_N$.

To derive the massless spectra of orientifolds we will work in light-cone
gauge. For example, in the closed untwisted sector the NS massless states are
$\psi_{-\oh}^\mu \vac$ which is invariant under $\th$, and 
$\psi_{-\oh}^i \vac$
which transforms as
\beq
\th^k  \psi_{-\oh}^i \vac =  {\rm e}^{2i\pi kv_i} \psi_{-\oh}^i \vac
\label{thpsi}
\eeq
Complex conjugates $\psi_{-\oh}^{-i}$ transform with a phase
${\rm e}^{-2i\pi kv_i}$.
The untwisted massless Ramond states are of the form
$| s_0 s_1 s_2 s_3 \rangle$
with $s_0, s_i = \pm \oh$ and odd number of minus signs to implement the
GSO projection. These states transform as
\beq
\theta^k |s_0 s_1 s_2 s_3 \rangle = {\rm e}^{2i\pi kv\cdot s}
|s_0 s_1 s_2 s_3 \rangle
\label{twistR}
\eeq
The condition $\pm v_1 \pm v_2 \pm v_3= 0$ ensures that there is a
gravitino in both the NS-R and R-NS type IIB untwisted sectors.
Projecting under $\Om$ then leads to \Neq1, \Deq4 supersymmetry.

Although type IIB is a theory of closed strings,
the orientifold projection requires both closed and open string sectors
for consistency.
The presence of the closed sector is clear. Its content is obtained
by retaining only those states which are
invariant under the orientifold group action and by including twisted
sectors. Details and examples are discussed in the next section.

The need for open string sectors can be justified in different ways.
An operative way of identifying them is to compute the partition function
(or generically, scattering amplitudes) in the closed sector Klein bottle
unoriented surface. Tadpole divergences are found. To cancel these tadpoles
and render the theory consistent, new contributions must be included
\cite{pcai}.
Introducing open strings leads to the required
cancellation for a specific structure of Chan-Paton charges.
Recall also that open strings are expected since when type IIB string
coordinates ends are identified up to the action of $\Omega$,
namely, $\Omega X(\sigma)=  X(\sigma)$, the mode
expansion of an open string is obtained.

The modern version of the above picture relies on the identification of
the tadpoles as non-cancelled charges. Orientifold fixed planes are
sources for $(p+1)$-forms originated in the Ramond-Ramond (R-R) sector.
Charge cancellation can be generically achieved by including  the right
number of D$p$-branes,  carrying opposite charge with respect to these
forms \cite{poldb}.
Open strings  have one end, labeled by $a$, on a D$p$-brane and
the other end, labeled by $b$, on a D$q$-brane . They give rise to
$pq$ string sectors. The $a,b$ labels correspond to the Chan-Paton factors
at each end of the string. We will construct
models with D9, D5 and D7-branes.

In some cases no tadpoles are present in the Klein bottle
amplitude and therefore there is no need for open string sectors.
As we explain in section~\ref{tadsec} and the appendix, there are on the
other hand cases cases where even
the inclusion of open string sectors is not enough to achieve tadpole
cancellation and the orientifolds are thus inconsistent.

\subsection{Closed string sector}
The spectrum in the closed sector of the orientifold is obtained from
those type IIB orbifold states invariant under $\Omega $ parity
transformations.
Orbifold states are constructed by coupling left and right
moving states of equal chirality, invariant under the orbifold group action.

The massless left NS states correspond to $\psi_{-\oh}^{\mu}|0\rangle $
vectors and to  matter scalars  $\psi_{-\oh}^{\pm i} |0\rangle$ ($i,j=1,2,3$).
Vectors are invariant under the orbifold twist $\theta$ action (\ref{twist})
while scalars acquire a phase ${\rm e}^{\pm 2\pi i v_i}$.
Right movers  are obtained by replacing $\psi \rightarrow {\tilde {\psi }}$.

By coupling left and right helicity $\pm \bf 1$
(under Lorentz little group $SO(2)$) vectors,
the graviton, an axion
(remnant of the \Deq10 NS-NS antisymmetric tensor) and a dilaton are found.
Since parity projection
keeps only symmetric combinations, only graviton ($\bf {\pm 2}$) and dilaton
multiplets are present in the orientifold.

The number of matter states depends on the type of twist $(v_1,v_2,v_3)$
under consideration. $\theta $ invariance requires
$\pm v_i \mp v_j= {\rm integer}$. We may rephrase this condition as
\begin{equation}
(r- {\tilde r})\cdot v= {\rm integer}
\label{uproy}
\end{equation}
where $r,{\tilde r}= (0,{\underline {\pm 1 ,0,0}})$ are $SO(8)$ vector
weights corresponding to the bosonized world-sheet fermions.
As an example consider $v= \frac1N (1,1,-2)$.
In this case, there are ten massless scalars,
\begin{equation}
 {\psi^{\pm i}_{-\oh}} |0\rangle _L  \otimes {\tilde {\psi}^{\pm j}_{-\oh}}
|0\rangle _R \\
\end{equation}
coming from $i=j=1,2,3$, $i=1, j=2$ and $i=2, j=1$. This
completes the  NS-NS sector of type IIB $Z_6$ orbifold for instance while
eight extra states ($i=1,2 ;j=3$) are present for $Z_3$ case.
 They combine into  five and nine  $\Omega $ invariant states respectively.

Massless Ramond left states, $|s_0 s_1 s_2 s_3 \rangle $
($s_0, s_i = \pm \oh$) carrying $\pm \oh$ helicity,  transform as indicated
in (\ref{twistR}). Right Ramond states transform similarly.
When both sector are coupled, twelve (twenty)  massless scalars survive the
orbifold projection in the $Z_6$ ($Z_3$) example above.
 Antisymmetric combinations lead thus to six (ten) $\Omega $ invariant states.
The twist invariant state with $s_1=s_2=s_3$ (and similarly for $\tilde s_i $)
(which is left from \Deq10 R-R antisymmetric tensor) will combine with the
NS-NS dilaton into a dilaton chiral multiplet $S$.
 The residual states combine into
five (nine) chiral massless states.

Construction of the R-NS sector proceeds in the same manner.
Supersymmetric partners of the above NS-NS and R-R bosons are found.
States invariant under $\Omega$ are obtained by taking the symmetrized
combinations R-NS + NS-R.

We have explicitly shown how to build up states in the untwisted sector. The
full one-loop torus amplitude can also be easily constructed.
For type IIB, we just couple the right movers sector (given for instance in
\cite{orbi}) with an identical (conjugate) expression for left
movers. In general, the trace over states can be written as
\begin{equation}
\cz = \frac1{N} \, \sum_{n,k=0}^{N-1}\cz_{\cal T}(\th^n, \th^k)
\end{equation}
The sum over $n$ is over twisted sectors whereas the sum over $k$ implements
the orbifold projection.
The first terms in a $q,\bar q$ expansion, ($q={\rm e}^{-2\pi t}$) read
\begin{eqnarray}
\cz_{\cal T}(\th^n,\th^k) \ \ \sim
& {}  & \tilde\chi (\th^n,\th^k)
\sum_r {[{\rm e}^{2i\pi(r+nv)kv}
q^{\oh (r+nv)^2-\oh } q^{E_0} (1+\cdots)] }
\nonumber \\
& {} &  \sum_{\tilde r}{ [{\rm e}^{-2i\pi({\tilde r}+nv)kv}
{{\bar q}^{\oh ({\tilde r}+nv)^2-\oh} {\bar q}^{E_0}(1+\cdots)}]}
\label{pf}
\end{eqnarray}
where $E_0= \sum_i {\oh n |v_i| (1-n |v_i|)}$.
Here $r$ and $\tilde r$ are $SO(8)$ weights and $\sum_{i=0}^3 r_i= {\rm odd}$
implements the GSO projection (similarly for $\tilde r $).
$\tilde\chi(\th^n,\th^k)$
takes into account the fixed point degeneracy \cite{orbi}. In many cases
$\tilde\chi(\th^n,\th^k)$ is just the
number of points left simultaneously fixed by $\theta^n$ and $\theta^k$, 
also $\tilde\chi(1,\theta^k)=1 $.

{} From eq.~(\ref{pf}) it follows that
massless states in the $\th^n$-twisted sector are given by $r,\tilde r$
such that
\begin{equation}
\oh (r+nv)^2 - \oh+{E_0}
= \oh ({\tilde r}+nv)^2-\oh+E_0 = 0
\label{mass}
\end{equation}
In the NS-NS sector both $r$ and $\tilde r$ are vector weights, whereas in
the R-R sector they are spinorial weights. In NS-R and R-NS we suitably
combine weights of both types.
{}From  eq.~(\ref{pf}) we can read the multiplicity
$D(n)$ of  massless states in the $\theta^n$ twisted sector, namely
\begin{equation}
D(n)= \frac1N \sum _{k=0}^{N-1} {\tilde\chi(\th^n ,\th^k)
{\rm e}^{2i\pi (r-\tilde r)\cdot kv}}
\label{mulc}
\end{equation}
For $n=0$ we recover eq.~(\ref{uproy}) above.

Only symmetric (antisymmetric) combinations are to be kept in the
NS-NS (R-R) sectors in the orientifold. Thus, (\ref{mulc}) gives the number
of chiral massless states in  the type IIB orientifold. Including all
twisted sectors gives the multiplicities shown in Table~\ref{clspec}.
We have included in the table only the $Z_N$, $Z_N\times Z_M$
orientifolds which are free of tadpoles and have at most one
sector of 5-branes (see below).

The generalization to $Z_N\times Z_M$ orbifolds \cite{zwart}
is straightforward. Denote by  $(\theta,\omega)$ the corresponding twists
with eigenvalues $(v_\th, v_\om)$. Then, the contribution
in a $\theta^n \omega^m$ twisted sector is obtained from the
above just by replacing $kv \rightarrow kv_\th+l v_\om$.
Recalling that a discrete torsion phase $\epsilon$ is now allowed \cite{fiq},
the degeneracy factor reads,
\begin{equation}
D(n,m)= \frac1{NM} \sum _{k=0}^{N-1} \sum _{l=0}^{M-1}{
\epsilon ^{(nk-ml)} \tilde\chi(\th^n\om^m, \th^k\om^l)
{\rm e}^{2i\pi (r-\tilde r)\cdot (kv_\th+lv_\om)}}
\label{mulct}
\end{equation}

\begin{table}[htb]
\footnotesize
\renewcommand{\arraystretch}{1.25}
\begin{center}
\begin{tabular}{|c|c|c|}
\hline
Twist Group   &  Untwisted moduli &  Twisted moduli  \\
\hline\hline
$Z_3 $ & $ 9 $ & 27  \\
\hline
$Z_3\times Z_3 $  & 3 & 81  \\
\hline
$Z_7$ & 3  &  21  \\
\hline
$Z_6$ & 5 &  29\\
\hline
$Z_6^\prime$ & 4 & 42 \\
\hline
$Z_3\times Z_6$ &3 & 71  \\
\hline
$Z_{12}  $ & 3 & 25\\
\hline
\end{tabular}
\end{center}
\caption{Number of chiral multiplets in closed string sectors for
some $Z_N$ and $Z_N\times Z_M$, \Deq4, \Neq1 type IIB
orientifolds. A dilaton multiplet must be added in the untwisted sector.
\label{clspec} }
\end{table}

\subsection{ Open string sector}
\label{opensec}

Let us now move to the open string sector. The type of D-branes present in
this sector depends on the content of the orientifold group. This is
expected since there are orientifold planes charged under R-R
fields twisted by the orbifold action \cite{gp}.  When the identity is
in $G_2$, the orientifold group contains ${\Omega}$ as an element
and there will be D9-branes. Following ref.~\cite{gj} we refer to these as
type A orientifolds. Contrariwise, type B orientifolds are those in
which D9-branes are not needed.

D5-branes are present whenever the orientifold group contains an
action of the type ${\Omega} R_i$, where $R_i$
is an order two element acting on  the two complex directions transverse
to $Y_i$. The corresponding $5_i$-branes live in \Deq4 space-time plus the
complex torus with coordinate $Y_i$.
For instance the $Z_6$ orientifold with $v={\frac{1}6}(1,1,-2)$
has $5_3$-branes whereas $Z_2 \times Z_2$ has sets of
$5_1$, $5_2$ and $5_3$-branes \cite{bl}.
It follows that  $Z_3$, $Z_7$ and $Z_3 \times Z_3$ orientifolds do not
contain D5-branes.

{}From T-duality arguments D7-branes are expected whenever $\Omega$
acts jointly with a reflection ${\cal R}_i$ on {\it one} complex plane
and $(-1)^{F_L}$ that changes sign
of all Ramond left states \cite{dp3,bz}.
We postpone the treatment of this case to section 4.

Open string states are denoted by $|\Psi, ab \rangle $, where $\Psi$
refers to world-sheet degrees of freedom while the $a,b$ Chan-Paton
indices are associated to the string endpoints on D$p$-branes and
D$q$-branes.
These Chan-Paton labels must be contracted with a hermitian matrix
$\lambda _{ab}$.
The action of a group element $g \in G_1$ is given by
\beq
g:|\Psi, ab \rangle    \rightarrow  (\gamma _{g,p})_{aa'} |g.\Psi,a'b'
\rangle (\gamma _{g,q})_{b'b}^{-1}
\label{gstate}
\eeq
where $\g_{g,p}$ and $\g_{g,q}$ are unitary matrices associated to $g$.
The action of $\Om h$, $h \in G_2$, is instead given by
\beq
\Om h:|\Psi, ab \rangle    \rightarrow  (\g_{\Om h,p})_{aa'} |h.\Psi,b'a'
\rangle (\g_{\Om h,q})_{b'b}^{-1}
\label{ohstate}
\eeq
Consistently with group multiplication, $\g_{\Om g,p}$ can be defined as
\beq
\g_{\Om g,p} \stackrel{\rm def}{=} \g_{g,p} \g_{\Om,p}
\label{gamom}
\eeq
The matrices $\g_{\Om,p}$ and hence $\g_{\Om g,p}$ are unitary.

Consistency with the orientifold group multiplication law implies several
constraints on the $\g$ matrices since they must provide a representation
of the group up to a phase \cite{gp}. For example, consider the
$G_1=G_2=Z_N$ case.
To a twist $\th^k$  there corresponds $\g _{\th^k, p} \equiv \g_{k,p}$
and $\g _{\Om {\th^k}, p} \equiv \g_{\Om k,p}$.
It can be shown that necessarily $\g_{0,p} = 1$ \cite{gp}.
Also, without loss of generality we can choose
\begin{equation}
\g_{k,p}=\g_{1,p}^k
\label{gamk}
\end{equation}
Since $\theta^N=1$, eq.~(\ref{gstate}) leads to
\begin{equation}
\gamma_{1,p}^N=\pm1
\label{gton}
\end{equation}
Similarly, from $\Omega ^2=1$ it follows that
\beq
\g_{\Om,p }=\pm\g^T_{\Om,p }
\label{gomt}
\eeq
Or in general, from $(\Omega \th^k)^2=\th^{2k}$,
\beq
\g_{\Om k,p}=\pm \g_{2k,p} \, \g^T_{\Om k,p}
\label{orcon1}
\eeq
Now, using eqs.~(\ref{gamom}), (\ref{gamk}), (\ref{orcon1}) and the
unitarity of the $\g$ matrices we obtain
\beq
\g_{k,p}^* = \pm \g_{\Om,p}^* \, \g_{k,p} \g_{\Om,p}
\label{afam}
\eeq
When there are different types of branes it is also necessary to consider
the action of $(\Omega \th^k)^2$ on $pq$ states.
In chapter 3 we will mostly consider
the same action of $\Om$ analyzed by Gimon and Polchinski (GP) \cite {gp}.
The GP action is such that $\Om^2 = (\pm i)^{(9-p)/2}$ on $9p$ states.
In particular, $\Om^2=-1$ on 95 states, implying that in eqs.~(\ref{orcon1})
and (\ref{afam}) there are opposite signs for 9 and 5-branes.

Cancellation of tadpoles imposes further conditions on the $\g$ matrices.
For instance it requires $\g_{\Om,9}^T = \g_{\Om,9}$. Hence, for 9-branes
we must take the plus sign in eqs.~(\ref{orcon1}) and (\ref{afam}). Since
we can choose $\g_{\Om,9}$ real we then have the condition
\beq
\g_{k,9}^* =  \g_{\Om,9} \, \g_{k,9} \g_{\Om,9}
\label{fam9}
\eeq
Now, for 5-branes the GP action implies $\g_{\Om,5}^T = -\g_{\Om,5}$.
Hence, for 5-branes we must take the minus sign in eqs.~(\ref{orcon1})
and (\ref{afam}). Since we can choose $\g_{\Om,5}$ pure imaginary
we then have the condition
\beq
\g_{k,5}^* =  \g_{\Om,5} \, \g_{k,5} \g_{\Om,5}
\label{fam5}
\eeq
The $\g_{k,p}$ matrices are determined from cancellation of `twisted'
tadpoles as discussed in next section. It turns out that they can
always be chosen diagonal.

The open string spectrum can be computed once the $\g$ matrices are found.
According to the string endpoints there are various $pq$ sectors and
moreover, D$p$-branes with $p < 9$ can sit at different fixed or non-fixed
points.
Here we will concentrate on models containing 9 and 5-branes, with all the
latter located on the particular fixed point corresponding to the
{\it origin} in the compact (transverse) space. This is an important case
since in this configuration
one gets maximal gauge symmetry. Furthermore, verification of tadpole
cancellation is much simpler. Different distributions of
5-branes on the various fixed points have to be analyzed case by case
since tadpole cancellation conditions might imply that there must always
be a number of residual 5-branes sitting at the origin which cannot move
to other fixed points. We will discuss this in some specific examples below.

We now describe the massless bosonic states in each $pq$ sector.
For the sake of clarity we restrict here to the orientifold
group generated by $\{G, \Omega \}$ with $G=Z_N$.
Generalization to $G=Z_N\times Z_M $ is straightforward. Extra
operations accompanying the action of $\Omega$ are considered in
specific examples.

\medskip\noindent
\underline{99-States}
\medskip

The massless NS states include gauge bosons
$\psi_{-\oh}^\mu|0,ab\rangle \, \lambda_{ab}^{(0)}$
and matter scalars $\psi_{-\oh}^i |0,ab \rangle \, \lambda_{ab}^{(i)}$.
The Chan-Paton matrices must be such that the full states are
invariant under the action of the orientifold group. Hence,
\beqa
\lambda^{(0)} & = & \gamma_{1,9}\lambda^{(0)}{\gamma^{-1}_{1,9}}  \
\quad\quad\quad ; \quad \quad
\lambda^{(0)} = -\gamma_{\Omega,9} {\lambda^{(0)}}^T {\gamma^{-1}_{\Omega,9}}
\nonumber \\[0.3ex]
\lambda^{(i)} & = & {\rm e}^{2\pi i v_i}\g_{1,9}\lambda^{(i)}\g^{-1}_{1,9}
\quad\quad ; \quad \quad
\lambda ^{(i)}  =  -\g_{\Omega,9}{\lambda^{(i)}}^T\g^{-1}_{\Omega,9}
\label{cons99}
\eeqa

\medskip\noindent
\underline{55-States}
\medskip

Massless NS states also include gauge bosons
$\psi_{-\oh}^\mu|0,ab\rangle \, \lambda_{ab}^{(0)}$
and matter scalars $\psi_{-\oh}^i |0,ab \rangle \, \lambda_{ab}^{(i)}$.
For $5_i$-branes at fixed points the Chan-Paton matrices must satisfy
\beqa
\lambda^{(0)} & = & \gamma_{1,5}\lambda^{(0)}{\gamma^{-1}_{1,5}} \
\quad\quad\quad ; \quad \quad
\lambda^{(0)} = -\gamma_{\Omega,5} {\lambda^{(0)}}^T {\gamma^{-1}_{\Omega,5}}
\nonumber \\[0.3ex]
\lambda^{(i)} & = & {\rm e}^{2\pi i v_i}\g_{1,5}\lambda^{(i)}\g^{-1}_{1,5}
\quad\quad ; \quad \quad
\lambda ^{(i)}  =  -\g_{\Omega,5}{\lambda^{(i)}}^T\g^{-1}_{\Omega,5}
\label{cons55}
\eeqa
But for $j \not= i$
\beq
\lambda^{(j)}  = e^{2\pi i v_j}\g_{1,5} \lambda^{(j)} {\g^{-1}_{1,5}}
\quad\quad ; \quad \quad
\lambda^{(j)}  =  \g_{\Omega,5} {\lambda^{(j)}}^T \g^{-1}_{\Omega,5}
\label{cons5j}
\eeq
The sign change in the $\Om$ projection is due to the DD boundary
conditions on the $j \not= i$ directions transverse to the $5_i$-branes.

\medskip\noindent
\underline{$5_i$ 9-States}
\medskip

For $5_i$-branes, coordinates orthogonal to $Y_i$ obey mixed DN boundary
conditions and have expansions with half-integer modded creation
operators. By world-sheet supersymmetry their fermionic partners in the
NS sector are integer modded. Their zero modes span a representation
of a Clifford algebra and can be labelled as $|s_j, s_k\rangle$, $j,k \ne i$,
with $s_j, s_k = \pm \oh$. The GSO projection further imposes
$s_j=s_k$. Under $\theta$, $|s_j, s_k\rangle$ picks up a phase
${\rm e}^{2\pi i( v_js_j+ v_ks_k)}$. Hence, the orientifold projection
on a state $|s_j, s_k, ab \rangle \lambda_{ab}$ implies
\begin{equation}
\lambda = e^{2\pi i( v_js_j+ v_ks_k)} \g_{1,5} \lambda  {\g^{-1}_{1,9}}
\label{cons59}
\end{equation}
Notice that here the index $a$ ($b$) lies on a 5-brane (9- brane).
$\Omega $ relates $59$  with $95$ sectors and does not impose extra
constraints on $\lambda$.

Interestingly enough, if  the $\lambda$ matrices are  recast
in a  Cartan-Weyl basis, constraints on Chan Paton matrices
emerge as  restrictions on weight  vectors. In this way
computing the spectrum becomes  greatly simplified.  Formally
it appears equivalent to the computation of the untwisted spectrum
of heterotic orbifolds.

For the sake of clarity let us first discuss how this is achieved in
the 99 sector.
We know that projecting under $\Omega $ parity, represented by a
symmetric $\gamma _{\Omega ,9}$ matrix in the 99 sector, gives the equation
$\lambda = -\gamma_{\Omega,9} {\lambda }^T {\gamma^{-1}_{\Omega,9}}$. The
original $32 \times 32$  unrestricted Chan-Paton (Hermitian) matrices are
therefore constrained to be $SO(32)$ generators. They can be organized
into charged generators $\lambda_a = E_a$, $a=1,\cdots, 480$, and Cartan
generators $\lambda_I = H_I$, $I=1,\cdots, 16$, such that
\begin{equation}
[H_I, E_a]=\rho_I^aE_a
\label{cw}
\end{equation}
where the $16$ dimensional vector with components $\rho_I^a$ is
the root associated to the generator $E_a$. These vectors are of
the form $(\underline {\pm 1, \pm 1,0,\cdots, 0})$, where underlining
indicates that all possible permutations must be considered. The
matrix $\gamma_{1,9}$ and its powers represent the action of the $Z_N$ group
on Chan Paton factors, and they correspond to elements of a  discrete
subgroup of the Abelian group spanned by the Cartan generators.
Hence, we can write
\begin{equation}
\gamma _{1,9}= e^{-2i\pi V \cdot H }
\label{Vdef}
\end{equation}
This equation defines the 16-dimensional `shift' vector $V_{(99)}=V$.
In section 2 we have seen  that $\gamma _{1,9}$ can be chosen
diagonal and furthermore $\gamma^N_{1,9} = \pm 1$.
Then, for example, when $\gamma^N=-1$ we will have a general structure
\beq
V= \frac1{2N}(1, \cdots, 1, 3,\cdots,3, 5,\cdots, 5, \cdots )
\label{spinsh}
\eeq
The number of $\frac{2l+1}{2N}$ entries is determined by
tadpole cancellation. Cartan generators are represented by
$2 \times 2$ $\sigma_3$ submatrices.  Explicit examples are
given in section 3.

 Recalling the formula
\begin{equation}
e^{-B} A e^B= \sum _{n=0} ^{\infty }
[A,B]_n
\end{equation}
with $[A,B]_{n+1}=[[A,B]_n,B]$, $[A,B]_{0}=A$,
and  using eq.~(\ref{cw}), it is easy to show that
\begin{equation}
\gamma_{1,9}E_a{\gamma^{-1}_{1,9}} =e^{-2i\pi \rho^a \cdot V}E_a
\end{equation}
Therefore we see from equation (\ref{cons99}) that gauge bosons
correspond to both, Cartan generators which trivially satisfy the
$\lambda ^{(0)}$ constraint, plus charged generators belonging
to a subset of
$SO(32)$ root vectors selected by
\begin{equation} \rho^a \cdot V_{(99)}= 0 {\rm \, mod \,} {\bf Z}
\label{v9p}
\end{equation}
Similarly, from the equation for $\lambda ^{(i)}$ in (\ref{cons99}) it
follows that matter states correspond to charged generators with
\begin{equation} \rho^a \cdot V_{(99)}= v_i {\rm \, mod \,} {\bf Z}
\label{m9p}
\end{equation}

Other open string sectors can be treated in a similar way.
For 5-branes on the bulk only $\Omega$ parity constraints are present
in eq.~(\ref{cons5j}). Since $\gamma _{\Omega,5}$ is antisymmetric
in this sector, symplectic group generators are obtained.
Furthermore, each  dynamical 5-brane contains $2N$ D5-branes
and the rank of the group is reduced.
The generators
 are associated to the same $SO$ root vectors given before plus long roots
$\underline {(\pm2,0,\cdots,0)}$.
For branes at fixed points we must impose the rest of the constraints in
eq.~(\ref{cons5j}) with $\g_{1,5}$ written in terms of an equivalent shift
$V_{(55)}$ as in (\ref{Vdef}). There is then a projection like (\ref{v9p}).
Moreover, since the equivalent shifts can be shown to be always of the form
(\ref{spinsh}), whenever 5-branes are at fixed points, long roots are
projected out.
If all 5-branes sit at the same fixed point we can take  $V_{(55)}= V_{(99)}$
and therefore, exactly the same spectrum
as in the 99 sector arises, now corresponding to multiplets
of the $SO(32)_{(55)}$ unitary subgroups of the 5-brane theory.

The $59$ sector is handled using an auxiliary
 $SO(64) \supset SO(32)_{(99)} \otimes SO(32)_{(55)}$ algebra.
Since we have generators acting simultaneously on both 9-branes and
5-branes  only roots of the form
\beq
W _{(95)}= W_{(9)}\otimes W_{(5)}=
({\underline {\pm 1, 0,  \cdots, 0}};{ \underline {\pm 1, 0, \cdots, 0}})
\label{w95def}
\eeq
must be considered. Here the first (second) $16$ components transform under
 $SO(32)_{(99)}$ ($SO(32)_{(55)}$).
 The shift in this sector is defined to be $V_{95}= V_{(99)}\otimes V_{(55)}$.
Using  (\ref{cons59}) we learn that massless states correspond
to $W_{(95)}$ roots satisfying
\begin{equation}
W _{(95)} \cdot V _{(95)}= (s_jv_j +s_kv_k) {\rm \, mod \,} {\bf Z}
\label{95sh}
\end{equation}
with $s_j=s_k=\pm \frac1{2}$, plus (minus) sign corresponding to particles
(antiparticles).

\subsection{Tadpole cancellation}
\label{tadsec}

In the orientifold theory the one-loop vacuum amplitudes include the torus,
the Klein bottle ($\ck$), the M\"obius strip ($\cam$) and the
cylinder ($\cc$). The last three have tadpole divergences from
exchange of massless states in the closed string channels.
By supersymmetry the total divergences vanish but consistency requires
separate cancellation of NS-NS and R-R tadpoles \cite{pcai}.
In refs.~\cite{sagnotti,bs,gp} it was shown how to extract these tadpoles
from the amplitudes in a $Z_2$ orientifold in \Deq6. These results
have been extended in both $D$=6,4 \cite{gj,dp2,kak3,zwart,odri}.
However, the general $Z_N$, \Deq4 orientifold tadpole cancellation
conditions have not been explicitly presented in those references.
In the appendix we give the general form of the amplitudes for
$T^6/\{ Z_N, \Om \}$ orientifolds in \Deq4 and comment briefly on extracting
the tadpoles. In this section we just describe and apply the main
results.

The $Z_N$ actions that can act crystallographically on a $T^6$ lattice and
lead to \Neq1 supersymmetry were classified long time ago \cite{dhvw}. The
list, with corresponding twist vectors, is given in table~\ref{tzn}.
Clearly, for even $N$, all the $Z_N$ have only one order two element
$R=\th^{N/2}$ that reflects $Y_1$ and $Y_2$. The corresponding orientifolds
will then have $5_3$-branes.

\begin{table}[htb]
\renewcommand{\arraystretch}{1.25}
\begin{center}
\begin{tabular}{|c|c||c|c||c|c|}
\hline
$Z_3$ & $\frac13(1,1,-2)$ & $Z_6^{\prime}$ & $\frac16(1,-3,2)$ &
$Z_8^{\prime}$ & $\frac18(1,-3,2)$ \\
$Z_4$ & $\frac14(1,1,-2)$ & $Z_7$ & $\frac17(1,2,-3)$ &
$Z_{12}$ & $\frac1{12}(1,-5,4)$ \\
$Z_6$ & $\frac16(1,1,-2)$ & $Z_8$ & $\frac18(1,3,-4)$ &
$Z_{12}^{\prime}$ & $\frac1{12}(1,5,-6)$ \\
\hline
\end{tabular}
\end{center}
\caption{$Z_N$ actions in \Deq4.}
\label{tzn}
\end{table}

The various tadpoles can be classified according to their volume
dependence. We denote by $V_i$, $i=1,2,3$, the volumes of the internal tori
and by $V_4$ the regulated space-time volume.
Also, as explained in the appendix, the $\ck$ amplitude has contributions
of type $\ck_1(\th^k)$ from untwisted closed strings, and,
for $N$ even, $\ck_R(\th^k)$ from $R$-twisted closed strings. The
$\cam$ amplitude receives contributions $\cam_p(\th^k)$ from open
strings with both ends on D$p$-branes.
The $\cc$ amplitude has pieces $\cc_{pq}(\th^k)$ from open strings with
ends on D$p$ and D$q$-branes.

In the $\ck$ amplitude, $\ck_1(1)$ has tadpoles proportional to
$V_4 V_1 V_2 V_3$ that can be cancelled by introducing $n_9$ 9-branes.
Taking into account the $\cam_9(1)$ and $\cc_{99}(1)$ divergences also
proportional to $V_4 V_1 V_2 V_3$, there is factorization and cancellation
of tadpoles provided that
\beq
\g_{\Om,9}^T = \g_{\Om,9}
\label{gom9t}
\eeq
and $n_9=32$. Recall that, as explained in section~\ref{opensec}, this
condition implies eq.~(\ref{fam9}). If we assume the
GP action of $\Omega$ we also obtain the eq.~(\ref{fam5}).

For $N$ odd the remaining divergences are all proportional to $V_4$
and factorize since we can always choose
\beq
\Tr (\gamma_{\Omega_k,9}^{-1} \, \gamma_{\Omega_k,9}^T) = \Tr \gamma_{2k,9}
\label{goch}
\eeq
Cancellation of twisted tadpoles then implies
\beq
\Tr \gamma_{2k,9} = 32 \, \prod_{j=1}^3  \cos \pi kv_j
\label{odd4}
\eeq
Also, we are free to choose $\g_{1,9}^N=1$ or $\g_{1,9}^N=-1$.

For $N$ even, $\ck_1(R)$ has tadpoles proportional to
$V_4 V_3/ V_1 V_2$ that can be cancelled by introducing $n_5$ $5_3$-branes.
Taking into account the $\cam_5(R)$ and $\cc_{55}(1)$ divergences also
proportional to $V_4 V_3 /V_1 V_2$ there is factorization and cancellation
of tadpoles provided that
\beq
\g_{\Om R,5}^T = \g_{\Om R,5}
\label{gom5t}
\eeq
and $n_5=32$.
If we assume the GP action of $\Om$ we then have
$\g_{\Om R,9}^T = -\g_{\Om R,9}$. In this case, using eqs.~(\ref{gamom})
and (\ref{fam9}), we immediately  find $\g_{R,9}^2 = -1$. Since $R=\th^{N/2}$
this then implies the condition $\g_{1,9}^N = -1$. Similarly,
$\g_{1,5}^N = -1$.

For $N$ even the type of other twisted tadpoles depends on the
specific form of $v$.
In all cases there are tadpoles proportional to $V_4 V_3$ that
arise from divergences in the amplitudes $\cc_{99}(R)$,
$\cc_{55}(R)$ and $\cc_{59}(R)$. These divergences factorize
into a square and give the condition
\beq
\Tr \g_{R,9} + 4 \Tr \g_{R,5,I} = 0
\label{zevent1}
\eeq
where $I=0,\cdots 15$, refers to the fixed points of $R=\th^{N/2}$.

Other tadpoles are most easily described case by case. In particular,
in $Z_4$, $Z_8$, $Z_8^{\prime}$ and
$Z_{12}^{\prime}$ we find that the Klein bottle amplitude has
divergences proportional to $V_4/V_3$ that cannot be cancelled against
any of the M\"obius strip or cylinder contributions.
Indeed, for instance in $Z_4$, if one tries to use $\g_{k,p}$ matrices
satisfying eq.~(\ref{zevent1}) together with
$\Tr \g_{k,9} + 2 \Tr \g_{k,5}=0$ from cancellation of tadpoles
proportional to $V_4$, one finds a massless spectrum that is not
free of gauge anomalies. This is a signal that there are left-over tadpoles.
This is a surprising result, since in \Deq6 one finds consistent
solutions, fully cancelling tadpoles, for all allowed $Z_N$.
Notice that in \Deq6 these $V_4/V_3$ tadpoles would be absent because
$V_3 \to \infty$.

For $Z_6$ there
are also tadpoles proportional to $V_4$ that imply the conditions
\beqa
\Tr \g_{k,9} & + &   \Tr \g_{k,5,0} \ \, = \ \, 0 \quad ; \quad k=1,5
\nonumber \\[0.2ex]
\Tr \g_{2,9} & + &  3\Tr \g_{2,5,0} \ \, = \ \, 16
\nonumber \\[0.2ex]
\Tr \g_{2,9} & + &  3\Tr \g_{2,5,J} \ \, = \ \, 4
\nonumber \\[0.2ex]
\Tr \g_{4,9} & + &  3\Tr \g_{4,5,0} \ \, = \ \, -16
\nonumber \\[0.2ex]
\Tr \g_{4,9} & + &  3\Tr \g_{4,5,J} \ \, = \ \, -4
\label{z6t2}
\eeqa
where $J=1,\cdots 8$, refers to the remaining fixed points
of $\th^2$ in the $(Y_1,Y_2)$ planes. In deriving these conditions
we have used eqs.~(\ref{goch}) and also
\beq
\Tr (\gamma_{\Omega_k,5}^{-1} \, \gamma_{\Omega_k,5}^T) = -\Tr \gamma_{2k,5}
\label{goch5}
\eeq
The minus sign in the right hand side is due to the GP action of $\Omega$.
Notice that as explained previously, $\g_{1,9}^6=-1$ and $\g_{1,5}^6=-1$.

In $Z_6^{\prime}$ the cylinder amplitudes for $\th, \th^5$
have tadpoles proportional to $V_4$. Cancellation gives the conditions
\beq
\Tr \g_{k,9}  -  2\Tr \g_{k,5,L} = 0 \quad ; \quad k=1,5
\label{z6pt2}
\eeq
where $L=0, \cdots, 3$, refers to the fixed points of $\th$ in the first
two tori.
In $\cc_{99}(\th^k)$, $\cam_9(\th^k)$ and $\ck_1(\th^k)$, $k=2,4$, there
are tadpoles proportional to $V_4 V_2$. Using eq.~(\ref{goch})
we find factorization leading to
\beq
\Tr \g_{2,9} = -8 \quad\quad ; \quad \quad \Tr \g_{4,9} = 8
\label{z6pt3}
\eeq
Finally, in $\cc_{55}(\th^k)$, $k=2,4$,
$\cam_5(\th^k)$ and $\ck_1(\th^k)$, $k=1,5$, there
are tadpoles proportional to $V_4/V_2$. Using eq.~(\ref{goch5})
we obtain
\beqa
& {} & \sum_M (\Tr g_{2,5,M})^2 + 16 \sum_L \Tr g_{2,5,L} = -64
\nonumber \\[0.3ex]
& {} & \sum_M (\Tr g_{4,5,M})^2 - 16 \sum_L \Tr g_{4,5,L} = -64
\label{z6pt4}
\eeqa
where $M=0, 1, 2$, refers to the fixed sets of $\th^2$ in the first two tori.
Notice that when all 5-branes sit at the origin (\ref{z6pt4}) gives
$\Tr \g_{2,5} = -8$ and $\Tr \g_{4,5} = 8$.
$Z_6^{\prime}$ can also be treated as $ Z_2 \times Z_3$ with twist
vectors $(\oh,-\oh,0)$ and $(-\frac13,0,\frac13)$. Applying the
general results of Zwart \cite{zwart} for $Z_N\times Z_M$ we
obtain tadpole cancellation conditions in complete agreement with
the above.

In $Z_{12}$ there are extra tadpoles proportional to $V_4 V_3$ since
$\th^3$ and $\th^9$ do not rotate the third direction. Cancellation of
these gives the conditions
\beq
\Tr \g_{k,9}  +  2\Tr \g_{k,5,L} = 0 \quad ; \quad k=3,9
\label{z12pt2}
\eeq
where $L$ refers to the fixed points of $\th^3$ in the first and second
complex coordinates. All other tadpoles are proportional to $V_4$ and
imply the constraints
\beqa
\Tr \g_{k,9} & - &   \Tr \g_{k,5,0} \ \, =
\ \, 0 \quad ; \quad k=1,2,5,7,10,11
\nonumber \\[0.2ex]
\Tr \g_{4,9} & + &  3\Tr \g_{4,5,0} \ \, = \ \, 16
\nonumber \\[0.2ex]
\Tr \g_{4,9} & + &  3\Tr \g_{4,5,J} \ \,  = \ \, 4
\nonumber \\[0.2ex]
\Tr \g_{8,9} & + &  3\Tr \g_{8,5,0} \ \, = \ \, -16
\nonumber \\[0.2ex]
\Tr \g_{8,9} & + &  3\Tr \g_{8,5,J} \ \, = \ \, -4
\label{z12t3}
\eeqa
where $J=1,\cdots 8$, refers to the remaining fixed points
of $\th^4$ in the $(Y_1,Y_2)$ planes.

\section{ $Z_N$ and $Z_N\times Z_M$  models with GP action}

In this section we study type IIB orientifolds based on
$T^6/\{ \Omega , G\}$ where $G$ denotes generators of a discrete
group $Z_N$ or $Z_N\times Z_M$, as discussed in the
previous chapter, leading to an unbroken \Neq1 supersymmetry in
four dimensions. Since $\Omega $ is  one of the generators of
the orientifold,  there are 9-branes in all of these
models. In addition, they contain as many independent sets
of 5-branes as different order two generators $G$ has.
Thus there will be models with zero, one or three
different sets of 5-branes.
Models in which $G$ contains a $Z_2\times Z_2$ subsector
will have in general three different sets of 5-branes.
The simplest $Z_2\times Z_2$ case was studied in \cite{bl} and has
a non-chiral spectra, a property shared by this subclass of models.

According to eq.~(\ref{fam9}), in all models
$\g_{\Om,9}$ is determined by requiring that
it be a real symmetric matrix that exchanges the eigenvalues of
$\g_{k,9}$ with their complex conjugates.
In this chapter we will assume the GP action of $\Omega $ so that
eq.~(\ref{fam5}) also holds. Thus,
$\g_{\Om,5}$ is determined by requiring that it be a pure
imaginary antisymmetric matrix that exchanges the eigenvalues of
$\g_{k,5}$ with their complex conjugates.
We have also seen that when $G$ contains a reflection $R$,
the GP action requires $\g_{R,9}^2 = -1$. Similarly, $\g_{R,5}^2 = -1$.
In next chapter we will discuss some aspects of different $\Omega $ actions.

Odd order orientifolds are particularly simple. Cancellation of untwisted
tadpoles requires 32 9-branes and there are no 5-branes since there is
no order two generator in $G$. The $Z_3$ case was
studied in refs.~\cite{ang, kak1} while $Z_7$ and $Z_3 \times Z_3$
were studied in \cite{kak2, kak3, zwart}.
We include them in our discussion for completeness.

We concentrate here on even order orientifolds with a single sector of 
5-branes.
We will first treat models in which all 5-branes sit on the particular
fixed point corresponding to the origin in the compact space.
Other distributions of 5-branes are analyzed in section \ref{wl5b}.
As explained in section \ref{tadsec} and the appendix, $Z_4$, $Z_8$,
$Z_8^{\prime}$ and $Z_{12}^{\prime}$ orientifolds have tadpoles that cannot
be cancelled by simply including 9 and 5-branes.
The only other orientifolds with one set of 5-branes are based on the
twists $Z_6$, $Z_6'$, $Z_{12}$ and $Z_3\times Z_6$.

We now consider each orientifold in more detail.

\subsection{ $Z_3$ }

With the choice $\g_\th^3=1$, cancellation of twisted tadpoles,
as given in eq.~(\ref{odd4}), requires
$\Tr \g_{\th}= -4 $. This fixes $\gamma _{\theta }$ uniquely:
\beq
\gamma _{\theta }  = \diag  (\alpha \id_{12}, \id_4,\alpha ^2 \id_{12}, \id_4 )
\label{z3sol}
\eeq
with $\alpha = {\rm e}^{2i\pi /3}$. Here and in the following $\id_r$ 
stands for
the $r\times r$ identity matrix. The open string spectrum can be easily
computed by using the auxiliary shift
\beq
V = {1\over 3} ( 1,1,1,1,1,1,1,1,1,1,1,1,0,0,0,0)
\eeq
We find a $U(12)\times SO(8)$ group and charged chiral fields
as shown in Table~\ref{specs}.
It is also possible to choose $\g_\th^3=-1$ leading to $\Tr \g_{\th}= 4$.
However, the resulting $\g_\th$ leads to the same group and spectrum.


\begin{table}[pht!]
\footnotesize
\renewcommand{\arraystretch}{1.25}
\begin{center}
\begin{tabular}{|c|c|c|}
\hline
Twist Group   & & \\
\cline{1-1}
Gauge Group & \raisebox{2.5ex}[0cm][0cm]{ (99)/(55) matter} &
\raisebox{2.5ex}[0cm][0cm]{ (95) matter}  \\
\hline\hline
$Z_3 $ & $3(12,8) + 3({\overline {66}},1) $  &  -  \\
\cline{1-1}
$U(12)\times SO(8)$ & & \\
\hline\hline
$Z_3\times Z_3 $  & $({\underline {4, 1,1}},8_v) +
({\underline {{\overline 4},{\overline 4},1 }},1)
+({\underline {6,1,1}},1)$ &   - \\
\cline{1-1}
$U(4)^3\times SO(8)$ &   & \\
\hline\hline
$Z_7$ &$({\underline {4, 1,1}},8_v) +({\underline {{\overline 4},
{\overline 4},1 }},1)
+({\underline {6,1,1}},1) $ &  -  \\
\cline{1-1}
$U(4)^3\times SO(8)$ &   $ + ({\ov 4},4,1,1)+(1,{\ov 4},4,1)+ (4,1,{\ov 4},1)$   & \\
\hline\hline
$Z_6$ &$2(15, 1, 1) +2(1,{\overline {15}},1)$
 &  $(6,1,1;6,1,1)+(1,{\overline 6}, 1; 1,{\overline 6},1)+$ \\
\cline{1-1}
      & $+2({\overline 6},  1, 4) + 2(1,6,\ov{4})$
 &      $ (1,6,1;1,1,\ov{4}) + (1,1,\ov{4};1,6,1)+ $  \\
$( U(6)^2\times U(4) )^2$    &$+ ({\ov 6},1,\ov{4})+(1,6,4)+(6,{\ov 6},1)$
& $({\overline 6},1, 1; 1,1,4)+ (1,1,4; {\ov 6},1,1)$ \\
\hline\hline
$Z_6'$ &$({\overline 4}, 1,8) +(1, 4,{\overline 8}) +({\underline {6,1 }},1)+
$ &  $({\overline 4},1,1;{\overline 4},1,1)+(1, 4, 1; 1,4,1)+$ \\
\cline{1-1}
 &   $(4,1,8)+(1,{\ov 4},{\ov 8}) +({\ov 4},4,1) + (1,1,28)$ &
 $(1,{\overline 4},1;1,1,8)+  (1,1,8;1,{\overline 4},1)+$ \\
 $(U(4)^2\times U(8) )^2$  & $ +(1,1, {\overline {28}})+
  (4,4,1)+({\overline 4},{\overline 4}, 1)$ & $ (4,1, 1; 1,1,{\overline 8})+
 (1,1,{\overline 8}; 4,1,1)$ \\
\hline\hline
$Z_3\times Z_6$ &$(2,2,1^5)+(1^2,2,2,1^3)+(1^4,2,2,1) +$
 &  $(2,1^6;1,2,1^5)+(1^2,2,1^4;1^3,2,1^3)+$  \\
\cline{1-1}
      &    $(1^4,2,1,4)+(1^5,2,{\overline 4})+(1,2,1^2,2,1^2)$
 &  $(1^4,2,1^2;1^5,2,1) + (1^5,2,1;1^6,{\overline 4})$  \\
 &   $ + (1^3,2,1,2,1) +  (2,1^5,4) + (2,1^4,2,1)  $
 &  $+ (1^4,2,1^2;1^6,4)$ \\
 &  $ + (1^2,2,1^3,{\overline 4}) + (1,2,1^4,{\overline 4}) + $
&  + same with groups reversed \\
$(U(2)^6\times U(4) )^2$  &  $ (1^2,2,1,2,1^2)  +(1^3,2,1^2,4) + 4(1^7)$ & \\
\hline\hline
$Z_{12}  $ &$({\underline { {\ov 3},1 }},{\underline { {\ov 3},1}},1,1)
+({\underline {3,1,1,1}},2,1) +$
 &  $({\ov 3},1^5;1,{\ov 3},1^4) + (1,3, 1^4; 1^5,2)+$   \\
\cline{1-1}
      &   $2(1,3,1,1,2,1)+ 2(3,1,1,1,1,2)+ $
 &      $ (3,1^5;1^4,2,1)+ (1^2,3,1^3;1^4,2,1) + $  \\
   &   $2(1,1,1,3,2,1)+2(1,1,3,1,1,2) +$ &
$(1^2,{\ov 3},1^3;1^3,{\ov 3},1^2) + (1^3,3,1^2;1^5,2)$ \\
$(U(3)^4\times U(2)^2 )^2$   &    $({\underline {3,1,1,1}},1,1) $
&  + same with groups reversed  \\
\hline\hline
\end{tabular}
\end{center}
\caption{
Gauge group and charged chiral multiplets in some $Z_N$ and $Z_N\times Z_M$,
\Deq4, \Neq1 type IIB
orientifolds with GP action. Only models with at most one set of 5-branes
are shown. All 5-branes sit on the  fixed point at the origin so that
in models with 5-branes the spectrum is explicitly  T-dual.
\label{specs} }
\end{table}


\subsection{ $Z_3\times Z_3$ }

The orbifold group is generated by twists $\theta, \omega $
whose action on the three complex coordinates are given by
$v_{\theta }= (\frac13,-\frac13,0)$ and $v_{\omega }=(0,\frac13,-\frac13)$.
Cancellation of twisted tadpoles requires \cite{kak3, zwart}
\beqa
\Tr \g_{\th} &=& \Tr \g_{\om} \ \, = \ \, \Tr \g_{\th \om} \ \, = \ \, 8
\nonumber \\[0.2ex]
\Tr \gamma_{\theta \omega^2} & = & - 4
\eeqa
Also, $\Tr \g_g^3=1$. Hence,
\beqa
\gamma _{\theta }  & =& \diag  (\alpha \id_4,\alpha ^2\id_4,
 \id_4,\id_4, \alpha ^2 \id_4,\alpha  \id_4, \id_4 ,
\id_4)
\nonumber \\[0.2ex]
\gamma _{\omega }  & =& \diag  (\id_4,
\alpha \id_4,\alpha ^2\id_4,
 \id_4,\id_4, \alpha ^2 \id_4,\alpha  \id_4 ,
\id_4)
\eeqa
with $\alpha = {\rm e}^{2i\pi/3}$.
The open string spectrum can be easily computed by using
the associated shift vectors:
\beqa
V_{\theta }&=& {1\over 3} (1,1,1,1,-1,-1,-1,-1,0,0,0,0,0,0,0,0)
\nonumber \\[0.2ex]
V_{\omega } & = & {1\over 3} (0,0,0,0,1,1,1,1,-1,-1,-1,-1,0,0,0,0)
\eeqa
The gauge group is $U(4)^3\times SO(8)$. The charged spectrum
is displayed in Table~\ref{specs}.

\subsection{ $Z_7$  }

The twist $\theta $ is generated by $v=\frac17 (1,2,-3) $.
Taking $\g_\th^7=1$, the twisted tadpole cancellation condition
eq.~(\ref{odd4}) implies
$\Tr \g_{\th}=32\cos\frac{\pi}7 \cos\frac{2\pi}7 \cos\frac{3\pi}7= 4$.
Then,
\beq
 \g_{\th} = \diag (\delta \id_4,\delta^2 \id_4, \bar\delta ^3\id_4,
 \id_4,\bar\delta  \id_4, \bar\delta^2  \id_4, \delta^3 \id_4, \id_4)
\eeq
where $\delta ={\rm e}^{2i\pi /7}$ and $\bar \delta = \delta^*$.
The open string spectrum can be computed using the associated shift
\beq
V = \frac17 (1,1,1,1,2,2,2,2,-3,-3,-3,-3,0,0,0,0)
\eeq
The gauge group is again $U(4)^3\times SO(8)$ although the charged spectrum is
slightly different from the $Z_3\times Z_3$ case (see Table 2).
It is also possible to choose $\g_\th^7=-1$ leading to $\Tr \g_{\th}= -4$.
However, the resulting $\g_\th$ leads to the same group and spectrum.

\subsection{ $Z_6$ }

The twist $\theta $ is generated by $v=\frac16(1,1,-2)$.
The twisted tadpole cancellation conditions are given in eqs.~(\ref{zevent1})
and (\ref{z6t2}). To simplify we consider the
case with maximal gauge symmetry in which all 32 5-branes sit at the origin
and we drop the fixed point subscript in $\g_{k,5}$.
This configuration is T-selfdual (under
duality transformations in the first two complex planes) and the gauge group
from 9-branes and 5-branes is the same. In this case tadpole cancellation
allows equal $\g_k$ matrices for 9-branes and 5-branes. Indeed, we find
\beqa
\Tr \gamma_{k,9} & = & \Tr \gamma_{k,5} \ \, = \ \, 0 \quad ; \quad k=1,3,5
\nonumber \\[0.2ex]
\Tr \gamma_{2, 9} & = & \Tr \gamma_{2,5} \ \, = \ \, 4
\nonumber \\[0.2ex]
\Tr \gamma_{4,9} & = & \Tr \gamma_{4,5} \ \, = \ \, -4
\eeqa
Also, condition (\ref{gom5t}) and the GP condition imply
$\g_{1,9}^6=-1$ and $\g_{1,5}^6=-1$ as we explained before.
The twist matrix is then \cite{kak1}
\beq
\gamma_{1,9}=\gamma_{1,5}=
\diag  ( \beta \id_6,\beta^5 \id_6, \beta^3\id_4,
\bar \beta \id_6, \bar\beta^5  \id_6, \bar\beta^3\id_4 )
\eeq
where $\beta = {\rm e}^{i\pi /6}$.
The matrices $\g_{\Om,9}$ and $\g_{\Om,5}$ are determined as
described before. We find
\beq
\g_{\Om,9} = \bmat{cc}0&\id_{16}\\ \id_{16} & 0 \emat
\quad ; \quad
\g_{\Om,5} = \bmat{cc}0&-i\id_{16}\\ i\id_{16} & 0 \emat
\label{goms95}
\eeq
$\g_{\Om,9}$ and $\g_{\Om,5}$ have this same form in
other even orientifolds in this section.

Computing the spectrum is substantially simplified using the shift
notation acting on $SO(32)$ roots. The associated shift corresponding to
$\g_{1,9}$ and $\g_{1,5}$ is
\beq
V_{(99)}=V_{(55)} = \frac1{12}(1,1,1,1,1,1,5,5,5,5,5,5,3,3,3,3)
\eeq
The gauge group in the (99) sector is given by $SO(32)$
roots $\rho$ verifying $\rho\cdot V_{(99)}=0$.
The (55) sector is identical so that the full group is
$(U(6)\times U(6)\times U(4))^2$.
Charged chiral matter fields in the (99) sector correspond
to roots verifying  $\rho \cdot V_{(99)}= \frac16, \frac16, -\frac13$
mod ${\bf Z}$ for each of the three compact complex planes.
The (55) sector has the same matter content. As explained in
section~\ref{opensec}, to find the massless chiral fields in the
$(95),(59)$ sectors we look for
weights $W_{(95)}$ verifying
$W_{(95)} \cdot V_{(95)}= \frac16 {\rm \, mod \,} {\bf Z} $.
In the end we obtain the spectrum displayed in Table~\ref{specs}.

\subsection{$Z_6^{\prime}$}

The twist $\theta $ has $v=\frac16(1,-3,2)$.
Tadpole cancellation conditions were given in section \ref{tadsec}.
With all 5-branes at the origin they imply
\beqa
\Tr \gamma_{k,9} & = & \Tr \gamma_{k,5}  \ \, = \ \, 0 \quad ; \quad k=1,3,5
\nonumber \\[0.2ex]
\Tr \gamma_{2, 9} & = & \Tr \gamma_{2,5}\ \, = \ \, -8
\nonumber \\[0.2ex]
\Tr \gamma_{4,9} & = & \Tr \gamma_{4,5} \ \, = \ \, 8
\label{z6ptrs}
\eeqa
We also have $\g_{1,9}^6=-1$ and $\g_{1,5}^6=-1$. Hence,
\beq
\gamma_{1,9}=\gamma_{1,5}=
\diag  ( \beta \id_4,\beta^5 \id_4, \beta^3\id_8,
\bar \beta \id_4, \bar\beta^5 \id_4, \bar\beta^3\id_8 )
\eeq
The associated shift acting on $SO(32)$ roots is
\beq
V = \frac1{12}(1,1,1,1,5,5,5,5,3,3,3,3,3,3,3,3)
\label{z6pV}
\eeq
both on the 9-brane and 5-brane sectors. The unbroken (99) or (55) gauge
group corresponds to $SO(32)$ roots $\rho $ verifying $\rho \cdot V=0$.
This yields $(U(4)\times U(4)\times U(8))^2$.
Charged chiral multiplets in the (99) or (55) sector are given by
roots satisfying  $\rho \cdot V =\frac16, \oh, \frac13$ mod ${\bf Z}$
respectively for each of the three compact complex planes.
To find the $(95),(59)$ chiral multiplets we
look for weights $W_{(95)}$
verifying $W_{(95)} \cdot V_{(95)}= -\frac16 {\rm \, mod \,} {\bf Z}$.
In this way we obtain the spectrum displayed in Table~\ref{specs}

\subsection{$Z_{12}$}

The twist $\theta $ is generated by $v={1\over 12}(1,-5,4)$.
Tadpole cancellation conditions were given in section \ref{tadsec}.
We choose a T-selfdual configuration in which all the 5-branes sit on
the fixed point at the origin. We then find
\beqa
\Tr \gamma_{k,9} & = & \Tr \gamma_{k,5}\ = \ 0  \quad ; \quad k\not= 4,8
\nonumber \\[0.2ex]
\Tr \gamma_{4,9} & = & \Tr \gamma_{4,5} \ = \ 4
\nonumber \\[0.2ex]
\Tr \gamma_{8,9} & = & \Tr \gamma_{8,5} \ = \ -4
\eeqa
Also, $\g_{1,9}^{12}=-1$ and $\g_{1,5}^{12}=-1$. The solution is then
\beq
\gamma_{1,9} = \gamma_{1,5}=
\diag  (\bar\zeta \id_3,\zeta^5\id_3, \bar\zeta^7 \id_3, \zeta^{11}\id_3,
 \zeta^3 \id_2, \zeta^9 \id_2,
\zeta \id_3,\bar\zeta^5\id_3, \zeta^7 \id_3,
 \bar\zeta^{11}\id_3  \bar\zeta^3 \id_2,  \bar\zeta^9 \id_2)
\eeq
where $\zeta = {\rm e}^{i\pi/12}$. Computing the massless spectrum with such
a matrix is really very cumbersome but becomes straightforward
using the shift notation. In this case
\beq
V_{(99)}=V_{(55)}=
{1\over {24} }(-1,-1,-1,5,5,5,-7,-7,-7,11,11,11,3,3,9,9)
\eeq
Proceeding as in the previous examples we obtain
gauge group $(U(3)^4\times U(2)^2)^2$ and charged chiral spectrum
as displayed in Table~\ref{specs}. One can check that
the $(99)$ and $(55)$ sectors have $SU(3)$ anomalies which are
appropriately cancelled by the chiral fields from the $(95)$ sector.

\subsection{$Z_3\times Z_6$}

The two generators $\theta $ and $\omega $ are
realized by $v_{\theta }=(\frac13, 0, -\frac13)$ and
$v_{\omega }=(\frac16,-\frac16,0)$. In this case $R=\omega^3$
leads to $5_3$-branes. Again we will treat the T-selfdual configuration
with all 5-branes at the origin. Applying the results of Zwart \cite{zwart}
we find that tadpole cancellation requires
\beqa
\Tr \gamma_{\theta,9 } & = &\Tr \gamma_{\theta, 5 } \ \, = \ \, 8
\nonumber \\[0.2ex]
\Tr \gamma_{\omega^2,9 } & = & \Tr \gamma_{\omega ^2,5 } \ \, = \ \, -8
\nonumber \\[0.2ex]
\Tr \gamma_{\th\omega^2,9 } & = & \Tr \gamma_{\th\omega^2,5 }\ \, = \ \, 4
\nonumber \\[0.2ex]
\Tr \gamma_{\omega^3,9} & = & \Tr \gamma_{\omega ^3,5} \ \, = \ \,
\Tr \gamma_{\omega,9 }\ \, = \ \,\Tr \gamma_{\omega,5 }\ \, = \ \,0
\nonumber \\[0.2ex]
\Tr \gamma_{\theta \omega,9 }& = &\Tr \gamma_{\theta \omega,5 }\ \, = \ \, 0
\eeqa
These constraints are fulfilled by the matrices
\beqa
\gamma_{\th,9 }& = &\gamma_{\th,5 } \ \, = \ \,
\diag  (\a^2\id_2, \id_2, \a\id_2, \id_2, \a\id_2, \a^2\id_2 , \id_4,
\a\id_2, \id_2, \a^2\id_2, \id_2, \a^2\id_2, \a\id_2 , \id_4)
\nonumber \\[0.2ex]
\gamma_{\om,9 }& = &\gamma_{\om,5 } \ \, = \ \,
\diag  ( \bar\beta \id_4, \bar\beta^5 \id_4, \beta^3\id_8,
 \beta   \id_4, \beta ^{5} \id_4, \bar\beta^3 \id_8)
\eeqa
where $\alpha={\rm e}^{2i\pi/3}$ and $\beta={\rm e}^{i\pi/6}$.
In this case it is particularly useful the use equivalent
shifts to compute the open string spectrum. These are
\beqa
V_{\theta,9 }& = & V_{\theta,5 } \ \, = \ \,
 {1\over {3}} (2,2,0,0,1,1,0,0,1,1,2,2,0,0,0,0)
\nonumber \\[0.2ex]
V_{\omega,9 } & = & V_{\omega,5 } \ \, = \ \,
{1\over {12}} (1,1,1,1,5,5,5,5,3,3,3,3,3,3,3,3)
\eeqa
The gauge group is $(U(4)\times U(2)^6)^2$. The charged particle
spectrum from the different sectors is shown in Table~\ref{specs}.
In this case the $SU(4)$ anomalies are separately
cancelled in each of the $(99)$, $(55)$ and $(95)$ sectors.

\subsection{Wilson lines and non-coincident 5-branes}
\label{wl5b}

In all the previous examples  we considered the most symmetric
situation in which there are no discrete nor continuous
Wilson lines and all 5-branes sit at the same fixed point
at the origin.
New models with different spectra and smaller gauge groups
can be obtained in the more general case in which
both possibilities (which in fact are T-dual to each other) are present.
We now discuss these possibilities and provide some examples.
We restrict to \Deq4, although the analysis applies equally well
to Wilson lines in \Deq6 orientifolds.

The orbifold action underlying the IIB orientifolds is
generated by the space group which involves elements $(\theta ,1)$, with
$\theta $  representing $Z_N$ rotations, and elements
$(1,e_m)$, with $e_m \in \Lambda$, $m=1,\cdots, 6$, where $T^6=R^6/\Lambda$.
The full space group is in general non-Abelian. The element $(\theta,1)$
is embedded in the open string sector through unitary
matrices $\gamma _{\theta , p}$, according to the D$p$-branes at
the endpoints.
In addition there can be background Wilson lines
which correspond to embeddings of the elements $(1,e_m)$ through matrices
$W_m$ into the 9-brane sector. To a fixed point $f$ of $\th^k$
there corresponds an element $(\theta^k, c_m e_m)$ such that
$(1-\th^k)f = c_m e_m$, for some integers $c_m$.
The 9-brane monodromy associated to this
fixed point will thus be $(\prod_m W_m^{c_m})\gamma_{k, 9}$.
The structure of the space group imposes constraints on $\g_{\th,9}$
and $W_m$. In particular, if $\theta$  rotates the lattice
vector $e_m$, $(\theta, e_m)^N=(1,0)$ and this in turn
implies $(W_m\gamma_{\theta, 9})^N=1$ (up to a
phase). If  $[W_m, \gamma_{\theta ,9}]=0$, then this
actually implies $W_m^N=1$ and we are dealing with a quantized
Wilson line. If $[W_m, \gamma_{\theta ,9}]\not=0$ the matrix
$W_m$ is in principle allowed to vary continuously and we
are dealing with a continuous Wilson line.

Let us consider fist the case of discrete Wilson lines.
Now there is not just one $\gamma _{\theta , 9}$ matrix
that must obey the tadpole cancellation conditions.
The different fixed points split into sets
feeling different gauge monodromies in the 9-brane
sector and tadpole conditions should apply to all
different embeddings $(\prod_m W_m^{c_m})\gamma_{k, 9}$.
This turns out to be a very stringent constraint.
Once the $W_m$ are determined we can compute the massless spectrum.
In the (99) sector we have to project with respect to the Wilson lines,
for both vector and chiral multiplets, according to
\begin{equation}
\lambda ^{(0)} =  W_m\lambda ^{(0)} W_m^{-1} \quad\quad ; \quad \quad
\lambda ^{(i)} =  W_m\lambda ^{(i)} W_m^{-1}
\label{wlproj}
\end{equation}
for all $m=1,\cdots, 6$.
In the $(95)$ sector when the 5-branes sit at
a fixed point in addition one has to take into account the precise
gauge monodromy corresponding to that specific fixed point.
In particular, if the point is fixed with respect to the
space group element $(\theta, c_me_m)$, in  eq.~(\ref{cons59})
one should replace $\gamma_{1, 9}$ by $(\prod_mW_m^{c_m})\gamma_{1, 9}$.

Consider as a first example the $Z_3$ orientifold discussed at
the beginning of this chapter. One can take for
the torus lattice $\Lambda$ the root lattice of $SU(3)^3$.
Consider the addition of a discrete
Wilson line $W_1$ along the first lattice vector $e_1$.
$W_1$ must be unitary and verify $W_1^3=1$.
Since $\theta e_1 = e_2$ , there must be also an identical
Wilson line $W_2=W_1$  along  $e_2$.
The three fixed points in the first lattice are the origin, $w_1$
with $(1-\th)w_1=e_1$ and $w_2$ with $(1-\th)w_2 = e_1+ e_2$.
Hence, the 27 fixed points
split into three sets of nine fixed points  feeling
monodromy $\gamma_{\theta , 9}$,  $W_1\gamma_{\theta , 9}$
and  $W_1^2\gamma_{\theta , 9}$ respectively. Tadpole
cancellation conditions will require
\begin{equation}
\Tr \gamma_{\theta , 9} =  \Tr W_1\gamma_{\theta , 9} =
\Tr W_1^2\gamma_{\theta , 9} = -4
\label{tcwl}
\end{equation}
We rewrite the twist (\ref{z3sol}) as
$\gamma_{\theta , 9}=
(\alpha \id_4, \alpha \id_4 , \alpha ^2 \id_4 , \id_4 ,
\alpha^2 \id_4 ,  \alpha ^2 \id_4 , \alpha \id_4 , \id_4 )$
in order to simplify calculations.
Then, the following choice for $W_1$ verifies the
constraints:
\begin{equation}
W_1  = \diag(\id_4,\alpha \id_4,\alpha \id_4,\alpha \id_4,
\id_4, \alpha ^2\id_4,\alpha ^2\id_4, \alpha ^2\id_4 )
\label{z3dwl}
 \end{equation} %
This Wilson line background breaks the $U(12)\times SO(8)$ gauge
symmetry down to $U(4)^4$.
The charged chiral multiplets transform as
\beq
3(1,{\overline 4},
4,1)+3(4,1,{\overline 4},1)+3({\overline 4},4,1,1)+ 3(1,1,1,6)
\label{finspec}
\eeq
This particular model was discussed in \cite{fin} and
has the peculiarity that the field theory associated to the first three
$SU(4)$ factors is finite. The scalar potential in this model has a flat
direction under which the $(99)$ chiral multiplets associated to one of the
three complex dimensions get a vev. As it is well known, this corresponds to
the addition of a continuous Wilson line. The $SU(4)^3$ gauge symmetry is
broken to the diagonal $SU(4)$ and in this particular example one gets a
model with \Neq4 global supersymmetry.

It is interesting to see how this diagonal group spectrum emerges when a
continuous Wilson  line ${\cal W}$ is turned on. Consider the
following proposal
\begin{equation}
{\cal W} =  \bmat{cccc}
\id_4 & 0 & 0 & 0 \\
0 &  W  & 0 & 0\\
0 & 0 & \id_4  & 0 \\
0 & 0 & 0 & W^*
\emat
\label{wlcont}
\end{equation}
where the $12 \times 12$ matrix $W$ is defined as
\begin{equation}
W=  \bmat{ccc}
w & a & a \\
a & w & a\\
a & a & w
\emat
\end{equation}
with $a$  an arbitrary complex number. It is easy to check that
constraints (\ref{tcwl}) are verified. Moreover,
$({\cal W}\gamma_{\theta, 9})^3=
\diag (\id_{12}, c \id_4, \id_{12}, c^* \id_4)$
where
$ c= (w^3-3a^2 w +2a^2)$.
For $a=0$ and $w= \alpha$ we recover the discrete Wilson line
(\ref{z3dwl}) discussed above. For $a\ne 0$
(and $\frac{w}a \ne 1,-2$) we can choose $a$ such that  $c=1$. Thus,
 we are left with a Wilson line completely rotated by the group action
 $\gamma_{\theta , 9}$ and depending on a complex continuous parameter
$w$.  When projections (\ref{wlproj}) are imposed,
$U(4)_{diag} \times U(4)$ gauge group and three chiral multiplets
$3(16,1)+3(1,6)$ are found.

It is illustrative to consider now an orientifold obtained
from $Z_3$ by a T-duality with respect to the first
two complex planes. In the notation of \cite{gj} this would
correspond to the \Deq4 analogue of the $Z_6^B$ orientifold.
Under this duality $\Omega  \rightarrow \Omega R $ and the orientifold
is generated by the order six element $\Omega \theta$, with $\theta $
generated
by $v=\frac16(1,1,-2)$. Thus, the orientifold group is
$\cg=\{ 1, \th^2 , \th^4 , \Omega R , \Omega \th, \Omega \th^5 \}$.
Since $\Omega $  is not an element
of the orientifold group, this model has no 9-branes. Cancellation
of untwisted tadpoles requires the presence of 32 5-branes.
Cancellation of twisted tadpoles further requires:
\beq
\Tr  \gamma_{k,5,0}= -4 \quad\quad   ; \quad\quad
\Tr  \gamma_{k, 5, J} = 0
\quad\quad ; \quad\quad k=2,4
 \label{z3b}
\eeq
where $J=1,...,8$, refers to fixed points in the first two complex
dimensions away from the origin ($J=0$ corresponds to the origin).
If all the 32 5-branes sit at the origin, the conditions
on $\gamma _{2,5,0}$ are analogous to those for 9-branes in the
T-dual model so that eq.~(\ref{z3sol}) would give a solution for
$\gamma _{2,5,0} $. This yields then gauge group $U(12)\times SO(8)$ and
chiral fields in $3(12,8)+3({\overline {66}},1)$. Now suppose that we
send some of the 5-branes to some other fixed point away from the origin.
Due to conditions (\ref{z3b}) we cannot send them all away from the origin,
a minimum of eight 5-branes must remain
with $\gamma _{2,5,0}=(\alpha \id_4 , \alpha ^2 \id_4 )$,
so that $\Tr \gamma _{2,5,0} = -4$. The other 24 5-branes
can leave the origin in groups of six 5-branes
(so that they form sets invariant under the orientifold action) and reach
some of the other fixed points. Take for example the case
in which 12 5-branes sit at the same
fixed point away from the origin.
The other 12 are related to the former by $\Omega R$ so
that projection under this generator does not give extra
constraints.
 Then, the tadpole condition on the 12 5-branes
has  solution
\beq
\gamma _{2,5,J }  = \diag  (\alpha \id_4 ,\alpha ^2 \id_4, \id_4 )
 \eeq
with $\alpha = {\rm e}^{2i\pi /3}$. The overall spectrum of this
5-brane configuration is as follows. The 8 5-branes at the
origin give gauge group $U(4)$ and three copies of chiral
fields in the antisymmetric representation. The other 24 5-branes yield
$U(4)^3$ and matter fields in
$3(1,{\overline 4}, 4)+3(4,1,{\overline 4})$$+3({\overline 4},4,1)$.
Thus, we recover exactly the same massless spectrum as the T-dual.
Now, if we send all the 24 5-branes to the bulk, they must travel
in $Z_3$ and $\Omega R$ invariant configurations. This means
that there are only four dynamical 5-branes leading to $SU(4)_{diag}$ with
three chiral multiplets in the adjoint. This corresponds to
giving non-vanishing vevs to the bi-fundamental fields present
in the configuration with 24 5-branes at the fixed point.

Let us now study Wilson lines in a model with
both 9-branes and  5-branes.
Consider the $Z_6'$ orientifold generated by the twist
$v=\frac16(1,-3,2)$. This can
be realized taking $\Lambda$ to be the $SU(3)\times SO(4) \times SU(3)$
root lattice. The properties of the
Wilson lines that can be added depend on what
complex direction the Wilson line wraps
around. Consider Wilson lines $W_3$, $W_4$,
wrapping around the second complex plane.
The condition $(\theta , e_{3,4})^6=1$
implies that  $W_{3,4}^2=1$. Suppose we add
a Wilson line $W_3$ around the direction $e_3$.
The four $\th$ fixed points in the first and second complex directions will
split into two  points with monodromy $\gamma_{\theta, 9}$ and
another two points with monodromy $W_3\gamma_{\theta ,9}$.
The three fixed points under $\theta^2$ will not
feel the Wilson line whereas the sixteen fixed points under
$R=\theta^3$ will split in two sets of eight fixed points each. In addition
to eqs.~(\ref{z6ptrs}), note that tadpole cancellation conditions will require
\begin{equation}
\Tr W_3\gamma_{3,9}= \Tr W_3\gamma_{1, 9}=0
\end{equation}
 Consider the following Wilson line
\begin{equation}
W_3 = \diag(\id_8,\id_r, -\id_{8-r},\id_8,\id_r, -\id_{8-r})
\end{equation}
This matrix verifies all the constraints.
The effect of $W_3$ is to break the gauge symmetry down to
$U(4)^2\times U(r)\times U(8-r)$. Consider the open string spectrum
in the particular $r=0$ simple case. The gauge group in the (99) sector
is as in the case without Wilson line, $U(4)^2\times U(8)$.
The chiral multiplets in this sector transform as
\beq
(1,1,28)+(1,1,{\overline {28}})+(4,4,1)+({\overline 4},{\overline 4},1)
+(6,1,1)+(1,6,1)+(4,{\overline 4},1)
\eeq
Concerning the chiral
multiplets in the (59) sector, if all 5-branes sit at the
origin, since the fixed point at the origin does not feel the
Wilson line, the spectrum is just that given in Table~\ref{specs}.
If all the 5-branes sit at one of the fixed points feeling the
Wilson line (those with coordinates $\oh e_3$ and $\oh (e_3+e_4)$ )
the (59) spectrum is still similar except for a flipping
$8\leftrightarrow {\overline 8}$ in the $U(8)$ of the 9-branes.
This is due to the fact remarked above that in the
projection one should replace $\gamma_{\theta , 9}$ by
$W_3\gamma_{\theta, 9}$.

Due to T-duality, in this $Z_6'$ orientifold
there must be an operation on the 5-branes degrees of freedom
which gives an analogous spectrum. In other words, certain
distributions of 5-branes on different fixed points  must
produce analogous physics. It is easy to find the configuration of
5-branes that reproduces the spectrum that we found for the
9-brane sector with the Wilson line $W_3$. We locate 16 of the
the 32 5-branes at the origin in the first two complex planes.
Those must fulfill the tadpole conditions in section~\ref{tadsec} that
have solution $\g_{1, 5,0}=(\beta \id_4 , \beta ^5 \id_4 ,
{\bar {\beta  }} \id_4 , {\bar {\beta }}^5 \id_4 )$.
These 16 5-branes give rise to $U(4)^2$ group
with charged fields transforming as $(4,4)+({\ov 4},{\ov 4})$
$+(6,1)+(1,6)$$+(4,{\overline 4})$.  Now we locate the
remaing 16 5-branes at one of the other three fixed points $L=1,2,3$,
in the first two complex planes. The
choice $\gamma _{1,5,L}=(\beta ^3 \id_8 , {\bar {\beta }}^3 \id_8 )$
is consistent with tadpole cancellation. It gives rise to
gauge group $U(8)$ and matter fields in $28+{\overline {28}}$. Notice that
the overall spectrum is the same as that obtained in the 9-brane sector
with the addition of the Wilson line $W_3$. Notice also that
there are no fields which transform non-trivially with respect
to {\it both} the gauge group from the 5-branes at the origin
and the group from the 5-branes at a different fixed point.

\section{$D=4$ Orientifolds with alternative discrete torsion projections}
\label{alpro}

As we already remarked, in even order type IIB orientifolds
tadpole cancellation is consistent with alternative ways of
realizing the orientifold  projection. This has been known in \Deq6 for
some time. Shortly after ref.~\cite{gp} appeared,
another way of orientifolding was presented \cite{dp1}, in
which  D9-branes are absent.
On the other hand, refs.~\cite{bz, dp3} working in an F-theory framework,
suggested another consistent way of performing the
orientifold projection.
Compactifying F-Theory on a $(h_{21},h_{11})=(51,3)$ Calabi-Yau (CY) manifold,
and using Sen's identification of the
fiber inversion with $\Omega (-1)^{F_L} $ in IIB theory \cite{sen05150}, gives
an orientifold group $ \{ \Omega (-1)^{F_L}{\cal R}_3, \Omega
(-1)^{F_R} {\cal R}_4, {\cal R}_3{\cal R}_4, 1 \}$.
Here ${\cal R}_i$ is a $Z_2$ inversion of  coordinate $Y_i$.
Tadpole conditions require the presence of two sets, 7, $7^\prime$,
of 7-branes.  Furthermore, the action of $\Omega^2$ in the
7-7' sector changed from  -1 (as in the GP model), to +1.
together with this, the new model had extra twisted tensors and symmetric
$\g_R$ matrices (compared to antisymmetric in the GP case).
In fact the GP model, or rather its T-dual, can also be constructed
in terms of these same generators and two sets of 7-branes \cite{sen11186}.
Hence, it became clear that the sign of
$\Omega^2$ in the mixed 7-7' sector is just a choice
that can be made each time a $Z_2$ projection is realized.
This is thus analogous to the discrete torsion degree of
freedom already encountered in heterotic orbifolds \cite{vafadt,fiq}.
This connection with discrete torsion was in fact suggested in \cite{dp3,gopm}.

The choice of sign for $\Omega^2$ can be shown to be related
to the symmetry or antisymmetry of the matrix
$\gamma_R$ that realizes the order two orientifold twist
on the Chan-Paton  matrices. It was noted in \cite{polten} that the presence
of certain couplings of R-R scalars to open string vectors
required the constraints
\beq
\g_R = - \g_{\Om} \g_R^T \g_{\Om}^{-1}
\eeq
for the standard $\Omega ^2= -1$ projection and
\beq
\g_R = + \g_{\Om^\prime} \g_R ^T \g_{\Om^\prime}^{-1}
\eeq
for an alternative $\Omega^{\prime 2}=+1$.
In what follows we will construct \Deq4, \Neq1 orientifolds
realizing the alternative projection.

\subsection{$Z_2\times Z_2$ orientifolds and discrete torsion}
\label{z2dt}

The \Deq4 analogue of the GP orientifold is the
$Z_2 \times Z_2$ orientifold of \cite{bl}. In this model the standard
$\Omega ^2=-1$ GP action is consistently assumed in mixed subsectors.
We would like now to construct \Deq4 models with alternative
actions.

In fact, the existence of these new models was conjectured in ref.~\cite{gopm}
that constructed \Deq4 models in terms of F-theory compactified in 
particularly simple Calabi-Yau four-folds.
In a few examples, such as their (C,C) model,
the complete spectrum coming from F-theory computations is available.
This is an affordable four-fold calculation in F-theory, as it admits a
$T^{8}/(Z_2)^3$ orbifold point. Indeed, it is the analogue of
the six-dimensional F-theory compactification on the (51,3) CY threefold that
we mentioned above.
The structure of the $Z_2$ fiber degenerations over the base implies
$SO(8)^{12}$ gauge group, no charged chiral multiplets and fifty-five moduli.

In \Deq6 it can be shown that the usual blowing up procedure,
corresponding to deforming the K\"ahler class, produces the (51,3) model
with $SO(8)^8$ gauge group and 17 tensors \cite{gopm}.
This is the model in \cite{bz, dp3}. Instead,
to get the GP model, with just one tensor, we have to resolve the
singularities by complex structure deformations that lead to
F-theory on the $(h_{21},h_{11})=(3,243)$
CY. These facts provide new clues for the discrete torsion analogy.
In ref. \cite{gopm} it was conjectured
that the model in \cite{bl} (which
only uses GP-type projections) can be obtained by an F-theory
compactification on a CY four-fold resolved by complex structure deformations.
Then, it is expected that blowing up the F-theory (C,C) model
corresponds to an orientifold with a `complementary
$\Om$ action'. This is what we now describe.

The orientifold group is generated by
$\{1, {\cal R}_2{\cal R}_3, \Om (-1)^{F_L}{\cal R}_1, \Om
(-1)^{F_L}{\cal R}_2 \}$.
We expect three sets of 7-branes, associated to elements
\beq
\Om (-1)^{F_L} {\cal R}_i  \stackrel{\rm def}{=} \Om_i
\quad \rightarrow  \quad 7_i{\rm -branes}
\eeq
In principle there could also be D3-branes,
due to the $\Om (-1)^{F_L} {\cal R}_1{\cal R}_2{\cal R}_3$ element,
but its absence in the F-theory formulation of the
same model indicates that they are not
present. We will make use of what we know from the \Deq6 model in
\cite{dp3,bz}.
As in that case we can put eight $7_i$-branes at each of the four orientifold
fixed points generated by $\Om_i$. Then, the $\g$ matrices are written as
$8\times8$ matrices. Twisted tadpole cancellation requires
\beqa
\Tr \g_{{\cal R}_1{\cal R}_2,7_1} & + &
\Tr \g_{{\cal R}_1{\cal R}_2,7_2} \ \, = \ \, 0
\nonumber \\
\Tr \g_{{\cal R}_2{\cal R}_3,7_2} & + &\Tr \g_{{\cal R}_2{\cal R}_3,7_3} \ \, =
\ \, 0
\nonumber \\
\Tr \g_{{\cal R}_1{\cal R}_3,7_1} & + &\Tr \g_{{\cal R}_1{\cal R}_3,7_3} \ \, =
\ \, 0
\eeqa
where the relative + sign is due to $\Om^2_{7_i7_j} = +1$  \cite{dp3,bz}.
We can then take
\beq
\begin{tabular}{l|cccccc}
&$\g_{\Om_1}$&$\g_{\Om_2}$&$\g_{\Om_3}$&$\g_{{\cal
R}_1{\cal R}_2}$&$\g_{{\cal R}_2{\cal R}_3}$&$\g_{{\cal R}_1{\cal R}_3}$\\
\hline
$7_1$ & $\id$ & $\id$ & $\id$ & $\id$ & $\id$ & $\id$ \\
$7_2$ & $\id$ & $-\id$ & $\id$ & $-\id$ & $-\id$ & $\id$ \\
$7_3$ & $-\id$ & $\id$ & $\id$ & $-\id$ & $\id$ & $-\id$ \\
\end{tabular}
\label{gtab1}
\eeq
\noindent
The signs for $\g_{\Om_i }$ are such that  $\g_{\Om_i} \g_{\Om_j} =
\g_{{\cal R}_i{\cal R}_j}$. The closed sector,
as in \cite{gopm}, gives fifty-five moduli. In open sectors, we have to
project with the $\g $'s above. Each of the three $7_i7_i$ sectors gives gauge
group $SO(8)^4$. For example, in the $7_1$ case, the Chan-Paton matrices
satisfy
\beqa
\lambda  & = & - \g_{{\Om_1};{7_1}} \lambda^T \, \g^{-1}_{{\Om_1};{7_1}} \ \,
= \ -\lambda^T
\nonumber \\
\lambda & = & \g_{{\cal R}_1{\cal R}_2;7_1} \lambda \,
\g^{-1}_{{\cal R}_1{\cal R}_2;7_1} \ \, = \ \, \lambda
\nonumber \\
\lambda & = & \g_{{\cal R}_2{\cal R}_3;7_1} \lambda \,
\g^{-1}_{{\cal R}_2{\cal R}_3;7_1} \ \, = \ \, \lambda
\eeqa
so that $\lambda$ is orthogonal and gives $SO(8)$ at each orientifold
plane. For chiral multiplets, as in \Deq6, the fact that the monodromy is
trivial ($\g_{{\cal R}_i{\cal R}_j} = \pm 1 $) kills all matter. In the
same way, considering the phases due to group actions, we find that
mixed $7_i7_j$ sectors do not add any other massless state.

We have therefore recovered the spectrum of the (C,C) F-theory
compactification in \cite{gopm}, a `discrete torsion' version of the
orientifold of \cite{bl}. The matrices in (\ref{gtab1}) representing the
$Z_2$ actions clearly are all symmetric, contrary to the matrices
of ref.~\cite{bl} that are antisymmetric, as in the original GP model.
We can nevertheless think of a model in which one
$Z_2$ twist is realized in GP way and the other in the way of
\cite{bz,dp3}. This option was also suggested in ref.~\cite{gopm} in terms
of a four-fold ambiguity in defining the $Z_2 \times Z_2$ orientifold.
The construction is quite similar to the one before, we just substitute
one $Z_2$ action from the symmetric form by one of the antisymmetric matrices
defined in \cite{bl}. More concretely, consider
\beq
\begin{tabular}{c|cccccc}
&$\g_{\Om_1}$&$\g_{\Om_2}$&$\g_{\Om_3}$&$\g_{{\cal
R}_1{\cal R}_2}$&$\g_{{\cal R}_2{\cal R}_3}$&$\g_{{\cal R}_1{\cal R}_3}$\\
\hline
$7_1$ & $\id$ & $M$ & $\id$ & $M$ & $M$ & $\id$ \\
$7_2$ & $M$ & $\id$ & $M$ & $M$ & $M$ & $\id$ \\
$7_3$ & $\id$ & $M$ & $-\id$ & $M$ & $-M$ & $-\id$ \\
\end{tabular}
\label{cuatrob}
\eeq
where
\beq
M = \bmat{cccc}0&\id_{2}&0&0 \\ -\id_{2}&0&0&0 \\ 0&0&0& -\id_2
	\\0&0&\id_2&0 \emat
\eeq
It is important that the set of matrices in (\ref{cuatrob}) satisfy all
conditions imposed by group multiplication.
These conditions are the same as in
\cite{bl}, except that when $\Om^{2}_{7_{1}7_{3}}$ enters, it gives an
extra -1 due to the new projection. Meanwhile $\Om^2_{7_17_2}$
remains equal to -1. The choice in (\ref{cuatrob})
cancels twisted tadpoles in accordance with
\beqa
\Tr \g_{{\cal R}_1{\cal R}_2,7_1}& -& \Tr \g_{{\cal R}_1{\cal R}_2,7_2} \ \,
= \ \, 0
\nonumber \\
\Tr \g_{{\cal R}_2{\cal R}_3,7_2}& -& \Tr \g_{{\cal R}_2{\cal R}_3,7_3} \ \,
= \ \, 0
\nonumber \\
\Tr \g_{{\cal R}_1{\cal R}_3,7_1}& + &\Tr \g_{{\cal R}_1{\cal R}_3,7_3} \ \,
=\ \, 0
\eeqa
The closed spectrum is still the same. In the open sector, the projection
by $M$ breaks $SO(8) \rightarrow SU(4)$ \cite{sen11186}. The three
$7_i7_i$ sectors give together $SU(4)^{12}$. To compute the matter spectrum
we just have to recall the phases corresponding to each group action, with
$\Om \rightarrow \pm {1}$ for DD, NN boundary conditions and ${\cal R}_i$
$\rightarrow -1$ when acting on $Y_i$. We get two copies of
 $({\underline {6,1,1,1}})$ from
each $7_i$ sector. In mixed sectors, the ${\cal R}_i{\cal R}_j $ twist kills
all states. The total spectrum is anomaly free.
It would be interesting to obtain the F-theory version of
this model for comparison.

\subsection{A chiral $Z_6^\prime$ model without 9-branes}
\label{z6gon}

One can also extend to \Deq4 the projection presented in ref.~\cite{dp1}.
One can for example construct a $Z_2 \times Z_3$ orientifold
with group $\{1,R\} \times \{1,\om, \om^2\}
\times \{1,\Om {\cal S}\}$, where $R$ is generated by
$v_R=(0,\oh,-\oh)$ and $\om$ by  $v_\om=(\frac13,-\frac13,0)$
whereas ${\cal S}$ is the transformation
${\cal S} : (Y_2 , Y_3) \rightarrow ( -Y_2, -Y_3 + \oh )$.
In principle, the
group elements $\Om {\cal S}$ and $\Om R{\cal S}$ would give D5-branes and
D9-branes respectively. But, due to ${\cal S}$, the Klein bottle amplitude
associated to D9-branes is free of tadpoles \cite{dp1}. There are still
$ \ck$ tadpoles proportional to $V_4V_1/V_3V_2$ that can be cancelled by
introducing D$5_1$-branes. To solve the model, we will consider a
configuration in which 16 D5-branes are on top of the origin and the other
16 sit on its image under ${\cal S}$. As in ref.~\cite{dp1}, we can take
$\g_{\Om {\cal S}} = \id $,
$\g_{R}=(\id_8,-\id_8)$. Also,
\beq
\g_{\om } = \diag (\a^2\id_2, \a\id_2, \id_4, \a\id_2, \a^2 \id_2, \id_4)
\eeq
The constraints on Chan-Paton factors read:
\beq
\lambda^{(0)} = \g_R \lambda^{(0)} \g_{R}^{-1} \quad\quad ; \quad\quad
 \lambda^{(0)} = \g_{\om} \lambda^{(0)} \g_{\om}^{-1}
\eeq
for gauge vectors while for matter
\beqa
\lambda^{(1)} & = & \g_R \lambda^{(1)} \g_{R}^{-1} \quad \quad \quad ;
\quad \quad
 \lambda^{(1)} = \alpha \g_{\om} \lambda^{(1)} \g_{\om}^{-1}
 \nonumber \\
\lambda^{(2)} & = & - \g_R \lambda^{(2)} \g_{R}^{-1} \quad \quad ;
\quad \quad
 \lambda^{(2)} = \alpha^{*} \g_{\om} \lambda^{(2)} \g_{\om}^{-1}
 \nonumber \\
\lambda^{(3)} & = & - \g_R \lambda^{(3)} \g_{R}^{-1} \quad \quad ; \quad
\quad
 \lambda^{(3)} = \g_{\om} \lambda^{(3)} \g_{\om}^{-1}
\eeqa
where $\alpha = {\rm e}^{2i\pi /3}$.
{} From the  $(55)$ sector we obtain gauge group $(U(2)^2\times U(4))^2$
and charged matter fields transforming as
\beqa
\psi^1 \  & : &
(2,1,{\overline  4};1,1,1)+(1,2,4;1,1,1)+(2,2,1;1,1,1)
\nonumber \\
    &  \ +  & (1,1,1;2,1,{\overline 4})+(1,1,1;1,2,4)+(1,1,1;2,2,1)
\nonumber \\
\psi^2 \ & : & (1,2,1;1,1,{\overline 4})+(1,1,4;2,1,1)+(2,1,1;1,2,1)
\nonumber \\
    &  \ + & (1,1,{\overline 4};1,2,1)+(2,1,1;1,1, 4)
+(1,2,1;2,1,1)
\nonumber \\
\psi^3 \ & : & (1,1,4;1,1,{\overline 4})+(1,1,{\overline 4};1,1,4)
\nonumber \\
  & \ + & 2(2,1,1;2,1,1)+2(1,2,1;1,2,1)
\label{gonzalo}
\eeqa
The model is chiral and anomaly free.

\section{Heterotic duals of $D=4$, $N=1$ orientifolds}
\label{hetduals}

The ten-dimensional $SO(32)$ heterotic string is supposed to be
S-dual to type I strings. This fact is already suggested by the form
of the ten-dimensional effective Lagrangian. By
dimensional reduction one can obtain that the
mapping between the dilatons of both dual theories
in $D$ dimensions is \cite{ang,anton2}
\begin{equation}
\Phi _I\ =\ {{6-D}\over 4}\ \Phi _H \ -\
{{D-2}\over {16}}\ \log\det G_H^{(10-D)}
\label{dillower}
\end{equation}
where $\Phi_I$ $(\Phi_H)$
is the type I (heterotic) dilaton.
$G_H$ is the metric of the (10$-D$) compact dimensions
in the heterotic frame. Notice that for \Deq10 indeed a
strongly coupled heterotic string maps to a weakly interacting
type I string. However, for \Deq4 eq.~(\ref{dillower}) shows that
there might be a weak-weak coupling duality for regions of
moduli space. This means that if we have a \Deq4 type I vacuum and
a heterotic vacuum which are dual to each other, their
spectra has to exactly match at weak coupling. It is important
to realize however that {\it the weakly coupled dual
of a perturbative (type I or heterotic) \Deq4 model need not
be a perturbative string construction.}

In ten dimensions the gauge group in type I originates in
open strings ending on 9-branes whereas in $SO(32)$ heterotic
they originate in the left-handed bosonic sector with
16 coordinates compactified on the $Spin(32)/Z_2$ lattice. Thus, in
searching for the heterotic duals of given \Deq4, \Neq1
orientifolds the obvious idea is to consider
heterotic $Z_N$ or $Z_N\times Z_M$ orbifolds whose
gauge degrees of freedom and untwisted chiral
states match the orientifold spectrum. It turns out
that this is possible in all cases. The identification is
particularly obvious if we choose the Cartan-Weyl representation
for the twist matrices $\gamma_{\theta , 9}$, as we discussed in
section~\ref{opensec}, and associate a shift
vector $V_{(99)}$ to these matrices. Thus, the natural mapping between
(99) orientifold states and heterotic untwisted sector is
\beqa
{\rm Type\ I}  & \longleftrightarrow &   SO(32)\ {\rm Heterotic}
\nonumber \\[0.2ex]
\psi_{-\oh}^\mu  |0,ab\rangle \, \lambda^{(0)}_{9,ab} & \longleftrightarrow
& \psi_{-\oh}^\mu \vac_R \otimes |P^I\rangle_L\ , |\partial X^I \rangle_L
\quad I=1,\cdots, 16
\nonumber \\[0.2ex]
\psi_{-\oh}^i |0,ab\rangle \, \lambda^{(i)}_{9,ab} \quad i=1,2,3
& \longleftrightarrow  &
\psi_{-\oh}^i \vac_R \otimes |P^I\rangle_L\ , |\partial X^I \rangle_L
\quad i=1,2,3  \nonumber \\[0.2ex]
\gamma _{1,9} = \exp(-2i\pi V_{(99)}\cdot H )
& \longleftrightarrow  &  V_{het} = V_{(99)},\ NV_{het}\in \Gamma
\label{mapping}
\eeqa
where $\Gamma$ is the $Spin(32)/Z_2$ lattice and
$P\in \Gamma$ are the gauge quantized momenta
of the heterotic string. On the heterotic side the action of the
twist $\theta $ in the gauge degrees of freedom is embedded through the
shift $V_{het}^I$ and the massless states are obtained
by projection $P \cdot V_{het}={\rm integer}$ for the gauge group and
$P\cdot V_{het}= v_i {\rm \, mod}$ integer for the $i=1,2,3$ untwisted chiral
multiplets. One can trivially check that with the identification
$V_{het}=V_{(99)}$ indeed the untwisted heterotic spectrum
precisely matches the $(99)$ sector of the candidate dual orientifold.

For a heterotic orbifold to be perturbatively consistent certain
modular invariant constraints must be fulfilled.
In particular level matching imposes for a $Z_N$ twisted
sector the constraint $N(V^2_{het}-v^2)=$ even.
Now, one can check that only the shifts $V_{(99)}$
in section 3  corresponding to odd $N$ ($Z_3,Z_3\times Z_3, Z_7)$ obey the
modular invariant constraints, while none of the even order twists do.
Indeed, perturbative heterotic duals for these
three orientifolds were proposed in refs.~\cite{ang,kak1,kak2}.
However, it is easy to obtain new shifts
that produce the same untwisted spectrum and {\it are}
modular invariant
also for even $N$. It is enough to consider any of the
even order shifts in section 3 and do the replacement
\beq
V_{het}\ =\ V_{(99)} \ \ \longrightarrow  V_{het}\ =\
V_{(99)}\ -\ (0,0,0,\cdots ,0,1)
\label{ellos}
\eeq
This fact was already noticed in the \Deq6 case in ref.~\cite{afiuv}.
Indeed, consider the simplest $Z_2$, \Deq6 GP orientifold.
One can check that in this case $V_{(99)}=\frac14(1,1,...,1,1)$,
which is not modular invariant, but the twist
$V_{het}=\frac14(1,1,...,1,-3)$  does obey the perturbative
modular invariant constraints. The new $V_{het}$ gives the heterotic dual
of a particular configuration in the $Z_2$ orientifold in which
there are 2 D5-branes at each of the 16 fixed points of $Z_2$ in \Deq6.
It was argued in ref.~\cite{afiuv} that the first (non-modular
invariant) shift for the heterotic model
gives rise to the heterotic
dual of a different configuration of the $Z_2$ orientifold
in which e.g. all 32 5-branes sit at the same fixed point.
The same was found for the other \Deq6, \Neq1 orientifolds,
the duals of orientifolds with all 5-branes sitting at the origin
are (non-perturbative) heterotic models in which the
standard modular invariant constraints are violated. This is
just as well, since most orientifold models have extra gauge and matter
degrees of freedom coming from the 5-brane sector which
can only appear on the heterotic side at the non-perturbative level.
In \Deq6 these non-perturbative effects may be understood as due to small
instantons either in the bulk or located at fixed points. Non-perturbative
heterotic orbifolds of this type were constructed in ref.~\cite{afiuv}.

The \Deq4, \Neq1 orientifolds discussed in this paper appear
to behave in a similar way concerning type I-heterotic duality.
It seems that the heterotic duals of orientifolds with all
5-branes sitting at the origin are non-perturbative
\Deq4, \Neq1 heterotic orbifolds in which  the usual modular
invariance constrains are violated and non-perturbative
gauge groups and fields arise due to small instanton effects.
Although, as we mentioned above, for each of the
orientifolds of even order one can find a perturbative
heterotic candidate dual obeying modular invariance
constraints, one cannot match the full massless spectra with
 the orientifolds since, to start with, these perturbative vacua
are missing the extra degrees of freedom associated to the 5-branes.
Furthermore, these perturbative heterotic vacua have extra charged matter
fields in their twisted sectors which also are missing in their
orientifold counterparts. From this point of view the \Deq6, $Z_2$
GP model is exceptional since there is one configuration of the
5-branes (two in each of the 16 fixed points) which precisely
matches the {\it perturbative} $Z_2$ heterotic orbifold that we
mentioned above. With that configuration there is no gauge group left
from the 5-brane sector.
 In the case of generic \Deq4, \Neq1 orientifolds of even order
such privileged 5-brane configurations seem difficult to find, if they
exist at all.

To exemplify the above discussion let us consider
a candidate perturbative  heterotic dual
of the $Z_6'$ orientifold. The mapping in eq.~(\ref{mapping}) suggests
constructing a perturbative
heterotic orbifold with a shift $V_{het}=V_{(99)}$ as
given in eq.~(\ref{z6pV}) appropriately shifted as in (\ref{ellos}).
 As we said, the untwisted sector
precisely matches the (99) sector of the orientifold.
On the other hand, the twisted sectors have content
\beqa
\theta, \theta ^5 \ \ &:& \ 12(4,1,1)+12(1,{\overline {4}}, 1)
\nonumber \\
\theta ^2, \theta ^4 \ \ & : & 9(6,1,1)+9(1,6,1)+6({\overline {4}},4,1)
   +3(4, {\overline 4} , 1)+18(1,1,1)
\nonumber \\
\theta^3\ \  & : & 4[(1,1,8)+(1,1,{\ov 8})+({\ov 4}, 1,1) +
    (1,4,1)] +8[(4,1,1)+(1,{\overline 4}, 1)]
\label{rabadan}
\eeqa
One can check that the contribution to non-Abelian anomalies coming
from $\theta , \theta ^5$ sectors is cancelled by that coming
from the $\theta ^2$, $\theta ^4$ sectors. The contribution of the
$\theta ^3$ particles exactly cancels against the untwisted
sector. Notice that the content of the $\theta ^3$ twisted sector
is identical to that of the (59) sector of the $Z_6'$ orientifold,
except for the obvious fact that multiplicities coming from the number
of fixed points in the heterotic orbifolds are representations
with respect to the (55) gauge group
on the orientifold side.  Although this coincidence would
 suggest that this perturbative heterotic model could be dual to the
$Z_6'$ orientifold, there is nothing in the orientifold
resembling the spectrum of $\theta ^n, n=1,2,4,5$ sectors.
Rather, one would expect this model to be dual
to some {\it non-perturbative type I} vacuum which has solitonic states
reproducing those sectors (in the same way that $SO(32)$ spinorial
representations are expected to appear at the non-perturbative level in
type I strings).

Heterotic duals for some of the \Deq4, \Neq1
 orientifolds of section~\ref{alpro}
with alternative orientifold projections can however be proposed.
Consider first the $Z_2\times Z_2$ orientifold with gauge group
$SO(8)^{12}$ of section~\ref{z2dt} and the
corresponding heterotic $Z_2\times Z_2$ orbifold generated
by the actions $v_{\theta }  =\oh(1,-1,0)$,
 $v_{\omega }=\oh(0,1,-1)$ acting on
a square $SO(4)^3$ compactification lattice. Add quantized Wilson lines
$a_1$, $a_2$ along, say, the first two compact coordinates. If
$\theta$ ($\omega $) is embedded in the $Spin(32)/Z_2$ gauge lattice
as a shift vector $A$ ($B$) given by
\beqa
A\ =\ a_1 & = &{1\over 2}(1,1,1,1,1,1,1,1,0,0,0,0,0,0,0,0)
\nonumber \\
B\ =\ a_2  & = &{1\over 2} (0,0,0,0,1,1,1,1,1,1,1,1,0,0,0,0)
\label{eseocho}
\eeqa
the gauge group is broken to $SO(8)^4$.
Notice that these shifts violate the
perturbative modular invariance constraints. The presence of the
Wilson lines projects out any massless  untwisted
charged field. Now, consider the
$\theta $ twisted sector. Its 16 fixed points are split into four
groups of four fixed points each feeling respectively
the gauge monodromies $A, A+a_2, A+a_1+a_2, A+a_1 $. The first three
shifts here have length  2 and there are in principle no perturbative
massless states in those sectors. However, as discussed in
refs.~\cite{intri, afiuv}, small instantons sitting at a $Z_2$ singularity
with such monodromy  generate a tensor multiplet in \Deq6
which in turn gives rise to a massless chiral singlet when reduced to \Deq4.
Thus, there are 12 singlets from those 12 fixed points. The other
4 fixed points under $\theta$ have trivial monodromy
since $A=a_1$ and $A$ is of order two. As remarked in ref.~\cite{blumintri},
5-branes at a $Z_2$ singularity with trivial monodromy
originate an $SO(8)$ vector multiplet plus one tensor multiplet in \Deq6.
Projecting down to \Deq4 one thus expects
from the four fixed points gauge group $SO(8)^4$ and 4 singlets.
Now, from the sector twisted under $\theta \omega $ a similar
massless spectrum, an $SO(8)^4$ and 16 singlets, is expected.
Finally, the 16 fixed pints under $\omega $ do not feel the Wilson
lines, all feel monodromy given by $B$. Thus one only expects
16 extra chiral singlets in the massless spectrum.
Putting all the pieces together, one obtains
the expected $SO(8)^{12}$ gauge group and the correct number
of moduli singlets to reproduce the spectrum of the orientifold.

In the same way one can find the candidate heterotic dual
for the $Z_2\times Z_3$ orientifold in section~\ref{z6gon}. Here
$v_{\theta }=\oh(0,1,-1)$ and
 $v_{\omega }=\frac13(1,-1,0)$. We add a Wilson line
$a_6$ around the sixth coordinate and embed $\theta$ and $\omega$
through shifts $A,B$ as follows:
\beqa
A  &  =& {1\over 2} (1,1,1,1,1,1,1,1,0,0,0,0,0,0,0,0)
\nonumber \\
B\ &= & {1\over 3} (0,0,0,0,1,1,-1,-1,1,1,-1,-1,0,0,0,0)
\nonumber \\
a_6 & =& {1\over 4}(1,1,1,1,\cdots ,1,1,1)
\label{gonza}
\eeqa
Again these shifts do not verify the  perturbative
modular invariance constraint.
One can check that the untwisted sector reproduces the
(55) sector of the orientifold in section~\ref{z6gon}.
Unlike in the previous example, there is no twisted sector
with trivial monodromy that could cause
extra non-perturbative gauge factors. As discussed in ref.\cite{afiuv}
the twisted  subsectors of order two and three have associated shifts
$A,B$ which only lead to extra singlets from twisted
sectors upon reduction
to $D=4$.

\section{Effective action and anomalous $U(1)$'s in $D\!=\!4$ orientifolds}
\label{efacc}

In this section we wish to discuss some general features
of the effective low-energy supergravity action
of \Deq4, \Neq1 orientifolds. We will concentrate on the class
of type IIB orientifolds discussed in chapter 3. Let us first describe
a few general properties of the massless closed string sector of these
theories. As discussed in chapter 2, besides the
supergravity multiplet,  closed strings give rise to a number of
untwisted chiral moduli singlets $T_i$ and twisted singlets $M_{\alpha}$.
Here $i=1,2,3$ except for the $Z_3$ and $Z_6$ orientifolds that have
extra off-diagonal moduli. In addition only $Z_6^\prime$ has one complex
structure field $U_2$. However, we will consider only the diagonal
untwisted moduli $T_i$ and the  dilaton chiral singlet $S$.
The dependence of these four complex scalars on the radii $R_i$ of the
three compact dimensions can be extracted from the \Neq2 results
of \cite{ant, anton2}
\beq
S\ =\ e^{-\phi }R_1R_2R_3  + i\theta \quad\quad ; \quad\quad
T_i=e^{-\phi }{{R_i}\over
{R_jR_k}} + i\eta _i  \quad\quad (i\not= j\not= k)
\label{sttt}
\eeq
where $\phi$ is the four-dimensional dilaton and $\theta ,\eta _i$
are R-R scalars. Let us concentrate now on orientifolds which contain
9-branes and one sector of 5-branes located at the origin
(all even order orientifolds in Table~\ref{specs}).
The gauge group in these models
has the structure $G_9\times G_5$, where $G_9$ ($G_5$) originates
in the 9(5)-brane sector of the theory. There are charged matter fields
of three types, fields $C^9_i$, $i=1,2,3$, charged under
$G_9$ only; fields $C^5_i$, $i=1,2,3$, charged under $G_5$ only;
and fields $C^{95}$ charged under both $G_9$ and $G_5$.
It is interesting to see how the different massless chiral fields
transform under T-duality. Suppose that the worldvolume of the
5-branes sweep the four non-compact dimensions and the third complex
plane. We already remarked that the configuration with all 5-branes
at the origin is invariant under T-duality transformations in the
first two complex planes. With the above definitions for
$S$ and $T_i$ one thus finds
\begin{eqnarray}
R_1\leftrightarrow {{\alpha '}\over {R_1}} \ \ & ; &\ \
R_2\leftrightarrow {{\alpha '}\over {R_2}}
\nonumber \\[0.2ex]
S &\leftrightarrow  &T_3 \nonumber\\
T_{1,2}&\leftrightarrow  &T_{2,1} \nonumber \\
C^9_i &\leftrightarrow  &C^5_i \nonumber \\
C^{95}&\leftrightarrow & C^{95}
\label{duales}
\end{eqnarray}
Then, we observe that under T-duality the roles of $S$ and $T_3$
are exchanged. The gauge kinetic functions
$f_{9,5}$ dependence on the
$S,T_i$ fields can also be extracted from the \Neq2 case  \cite{ant}
\beq
f_9\ =\ S \quad\quad ; \quad\quad f_5\ =\ T_3
\label{efesi}
\eeq
This is consistent with the fact that under T-duality in the first two
complex planes $S$ goes to $T_3$ and the r\^oles of 5-branes and
9-branes are exchanged. Notice that from the heterotic point of
view the gauge interactions from 5-branes are non-perturbative
and this tells us that their strengths are governed
(in the dual heterotic) by the moduli rather than the dilaton.

In fact there are reasons to argue that both $f_9$ and $f_5$
also depend linearly on closed string twisted singlets $M_{\alpha}$.
One can easily check that all orientifolds in Table~\ref{specs}
have anomalous $U(1)_X$'s in their spectra. We know that in
\Deq4, \Neq1 heterotic vacua those anomalies are cancelled
by a four-dimensional version of the Green-Schwarz \cite{gs}
mechanism in which $\pim S$ transforms as $\pim S\rightarrow
\pim S- \delta _{GS}\Lambda (x) $ under a $U(1)_X$ gauge transformation with
gauge parameter $\Lambda (x)$. Since in the perturbative heterotic
$\pim S$ couples to $F{\tilde F }$
in a universal manner to all groups, the $U(1)_X$ anomaly can
cancel as long as the mixed anomaly of the $U(1)_X$ with all
gauge interactions are in the ratio of the Kac-Moody levels of the
gauge algebras. Furthermore, since there is only
one dilaton field $S$ to do the trick, in perturbative heterotic
vacua there is at most only one anomalous $U(1)_X$.
Equation (\ref{efesi}) already tells us that in the case of
type IIB, \Deq4 orientifolds there will be in general more than one
anomalous $U(1)_X$ since not only $S$ but moduli fields like
$T_3$ have suitable couplings for a Green-Schwarz mechanism to work.

More precisely, one finds that in the absence of Wilson lines
the orientifolds $Z_3$ and $Z_3\times Z_3$ have only
one anomalous $U(1)_X$, but several if Wilson lines are added. 
The $Z_7$ orientifold has three anomalous $U(1)_X$'s when there 
are no Wilson lines.
For models with both 5-branes and 9-branes one finds several
anomalous $U(1)_X$'s. For example, one finds three anomalous
$U(1)_X$'s in the $Z_6$ orientifold of chapter 3. In order for
all the anomalies to cancel, non-linear transformations of
the $S$ and $T_3$ fields are not enough. This is particularly
obvious in odd models like $Z_3$ in which there is no
5-brane sector and the $S$ field couples universally to
the $SU(12)$ and $SO(8)$ factors. One can easily show that
the mixed anomaly of the $U(1)_X$ with those two factors is
different and could not possibly be cancelled by a shift in
$\pim S$. What happens is that a linear combination of 27
twisted moduli fields $M_{\alpha }$, $\alpha =1,..,27$, do also
get shifted under a $U(1)_X$ transformation. In addition,
direct couplings of the $M_{\alpha }$ closed string states with
$F_9 \tilde F_9$ must exist for the mechanism to work. That this is the
case can also be confirmed by studying the heterotic dual
of this model.  Something analogous is expected in the
case of models with 5-branes. Thus, $U(1)_X$ anomaly
cancellation in this class of orientifolds requires
gauge kinetic functions of the form
\beq
f_9\ =\ S\ +\ \sum_{\alpha } c_{\alpha}^9 M_{\alpha }
\quad\quad ; \quad \quad
f_5\ =\ T_3\ +\ \sum_{\alpha } c_{\alpha}^5 M_{\alpha }
\label{corref}
\eeq
where $c_{\alpha }^{9,5}$ are constant coefficients.
Multiple $U(1)_X$ anomaly cancellation is achieved by
non-trivial transformations of $S,T_3$ and $M_{\alpha }$ under
the $U(1)_X$'s. Notice that this is analogous, though not identical,
to the generalized Green-Schwarz mechanism in \Deq6 theories
suggested in ref.~\cite{sagcan}.

Using symmetry arguments one can also extract some of the
relevant terms for the K\"ahler potential in
orientifolds with 9-branes and one set of 5-branes.
The equivalent terms for the \Neq2 case were obtained in ref.~\cite{ant}.
Considering the truncation from the \Deq10 type I
action and imposing invariance under T-duality with respect
to the transformations in eq.~(\ref{duales}) one gets
\beqa
K &=& -\log(S+S^* + |C_3^5|^2)   -  \log(T_3+T_3^* + |C_3^9|^2 )
\nonumber \\[0.2ex]
   & - & \log(T_1+T_1^*+|C_1^9|^2 +|C_2^5|^2 )  -
\log(T_2+T_2^*+|C_2^9|^2 +|C_1^5|^2 )
\nonumber\\[0.2ex]
  &+ & {{|C^{95}|^2}\over {(T_1+T_1^*)^{1/2}(T_2+T_2^*)^{1/2}}}
\label{kali}
\eeqa
Indeed the above formula is invariant under T-duality with respect to
the first two complex planes and with respect to the
Peccei-Quinn symmetries corresponding to shifts of the
imaginary parts of $S$ and $T_i$. The form of the metric of the
charged $C^{95}$ fields is suggested by  T-duality invariance and the
fact that the $(95)$ sector of these theories behave as a sort
of $Z_2$ twisted sector.
The perturbative trilinear superpotential $W$
has the structure
\beq
W\ =\ C_1^9C_2^9C_3^9\ +\ C_1^5C_2^5C_3^5\ +\
C^{95}C^{95}C_3^9 \ +\ C^{95}C^{95}C_3^5
\label{superp}
\eeq
It is also explicitly invariant under T-duality in the first two
complex planes. One can also test in specific examples that
gauge quantum numbers are consistent with the existence of these
couplings.

\bigskip

\bigskip

\bigskip

\centerline{\bf Acknowledgements}
We are grateful to A.~Uranga, J.~Maldacena and G. Zwart for discussions
and to
R.~Rabadan for computing the spectrum of the model in eq.~(\ref{rabadan}).
G.A. thanks the AS-ICTP for hospitality and financial support.
L.E.I. and G.V. thank CICYT (Spain) for financial support.
A.F. thanks CONICIT and CDCH-UCV for research grants, the AS-ICTP as
well as the CERN TH-Division for hospitality and support,
and Fundaci\'on Polar for travel aid.

\newpage
\section{Appendix}

{\large \bf 1-loop amplitudes and tadpoles}

\bigskip

We want to compute the tadpoles for $T^6/\{Z_N,\Om\}$ type IIB orientifolds. 
We start by writing down the Klein bottle amplitude given by
\beq
\ck = \frac{V_4}{8N} \sum_{n,k=0}^{N-1}  \int_0^\infty \, \frac{dt}t \,
(4\pi^2 \a^\prime t)^{-2} \, \cz_{\ck}(\th^n,\th^k)
\label{kamp}
\eeq
where
\beq
\cz_{\ck}(\th^n,\th^k) = \Tr \{(1+(-1)^F) \Om \, \th^k \,
{\rm e}^{-2\pi t[L_0(\th^n) + \ov{L}_0(\th^n)]} \}
\label{zkdef}
\eeq
The contribution of the uncompactified momenta is already extracted in
(\ref{kamp}).
Recall that $V_4$ denotes the regularized space-time volume.
Since $\Om$ exchanges $\th^n$ with $\th^{N-n}$, only $n=0$ and
$n=\frac{N}2$, if $N$ is even, do survive the trace.
$\cz_{\ck}(1,\th^k)$ and $\cz_{\ck}(R,\th^k)$ lead to pieces
$\ck_1(\th^k)$ and $\ck_R(\th^k)$ whose definition is obvious from
(\ref{kamp}).

The trace in $\cz_{\ck}$ can be evaluated in a standard way using
$\vt$ functions to write the contributions of complex bosons and fermions.
Also, the GSO projection is implemented by summing over spin structures.
Then, taking into account the insertion of $\Om$ we find
\beq
\cz_{\ck}(1,\th^k) = \sum_{\a,\b=0,\oh}  \eta_{\a,\b} \
\frac{\tilde\vt[{\a \atop \b}]}{\tilde\eta^3} \ \prod_{i=1}^3
\  \frac{\tilde\vt[{\a \atop {\b + 2kv_i}}]}{\tilde\eta}  \,
\frac{-2\sin 2\pi k v_i \,  \tilde\eta}{\tilde\vt[{\oh \atop {\oh +2kv_i}}]} \
\label{k1}
\eeq
where $\eta_{0,0} =-\eta_{0,\oh} = -\eta_{\oh,0} =1$.
The $\vt$ and $\eta$ functions are defined in the compendium at the end of
this appendix. The tilde indicates that the argument is
$\tilde q = q^2 = {\rm e}^{-4\pi t}$. This result is strictly valid only
if $2kv_i \not = {\rm integer}$. If $kv_i=$integer, (\ref{k1}) has a well
defined limit but we must also include a sum over quantized momenta in
the $Y_i$ direction. If $kv_i=$half-integer, (\ref{k1}) has again a well
defined limit but we must also include a sum over windings in the $Y_i$
direction. Upon taking the limit $t\to 0$, using the Poisson
resummation formula, a sum over quantized momenta in $Y_i$ gives an
internal volume factor $V_i$, whereas a sum over windings
in $Y_i$ gives a factor $1/V_i$.
For example, in $Z_6^\prime$ with $v=\frac16(1,-3,2)$ we have the
following overall volume dependence in $\ck_1(\th^k)$: $V_4V_1V_2V_3$ for 
$k=0$,
$V_4/V_2$ for $k=1,5$, $V_4V_2$ for $k=2,4$, and $V_4V_3/V_1V_2$ for $k=3$.

Let us now write $\cz_{\ck}(R,\th^k)$ that appears when $N$ is even.
We assume that $v$, as in Table~\ref{tzn}, has been chosen so that to
$R=\th^{N/2}$ there corresponds
$\frac{N}2 v = (\oh, \oh, 0)$ mod ${\bf Z}$. Then,
\beq
\cz_{\ck}(R,\th^k) =  \tilde \chi(\th^{N/2}, \th^k)  \sum_{\a,\b=0,\oh} \!
\eta_{\a,\b} \
\frac{\tilde\vt[{\a \atop \b}]}{\tilde\eta^3} \left ( \prod_{i=1}^2
\frac{\tilde\vt[{{\a+\oh} \atop {\b + 2kv_i}}]}
{\tilde\vt[{0 \atop {\oh +2kv_i}}]} \right ) \,
\frac{\tilde\vt[{\a \atop {\b + 2kv_3}}] \, (-2\sin 2\pi k v_3)}
{\tilde\vt[{\oh \atop {\oh +2kv_3}}]}
\label{kR}
\eeq
$\tilde \chi(\th^{N/2}, \th^k)$ is a factor that takes into
account the fixed point degeneracy \cite{orbi}.
Eq.~(\ref{kR}) is strictly valid only if $2kv_3 \not = {\rm integer}$.
If $kv_3=$integer, there is a well defined limit but we must also
include a sum over quantized momenta in $Y_3$. If $kv_3=$half-integer,
again there is a well defined limit but we must also include a sum over
windings in $Y_3$.

Both (\ref{k1}) and (\ref{kR}) vanish by virtue of supersymmetry. Indeed,
choosing, as in Table~\ref{tzn}, $v_1+v_2+v_3=0$, and using the
identities (\ref{abtru}), we find
\beqa
\cz_{\ck}(1,\th^k) & \! = \! &
(1-1)  \frac{\tilde\vartheta[{0 \atop \oh}]}{\tilde\eta^3}
\prod_{i=1}^3 (-2 \sin 2\pi kv_i)
\frac{\tilde\vt[{0 \atop {\oh + 2kv_i}}]}
{\tilde\vt[{\oh \atop {\oh + 2kv_i}}] }
\nonumber\\[0.3ex]
\cz_{\ck}(R,\th^k) & \! = \! & (1-1)  \tilde \chi(\th^{N/2}\! , \th^k)
\frac{\tilde\vartheta[{0 \atop \oh}]}{\tilde\eta^3}
\left ( \prod_{i=1}^2
\frac{\tilde\vt[{\oh \atop {\oh + 2kv_i}}]}{\tilde\vt[{0 \atop
{\oh + 2kv_i}}] } \right )
\frac{(-2 \sin 2\pi kv_3)\, \tilde\vt[{0 \atop {\oh + 2kv_3}}]}
{\tilde\vt[{\oh \atop {\oh + 2kv_3}}] }
\label{kids}
\eeqa
Notice that in $\cz_{\ck}(R,\th^k)$ the expression multiplying $(1-1)$
vanishes identically when $k=0,N/2$ and when $2kv_2={\rm integer}$ as in
$Z_6^\prime$.

The next step is to extract the divergences as $t\to 0$. To this end we
use the identities (\ref{modt}) given in the compendium. Roughly speaking,
we find $\cz_{\ck} \to (1-1)(2t)\times {\rm factors}$. More precisely,
take for example $\cz_{\ck}(1,\th^k)$ and for simplicity
assume $2kv_i \not= {\rm integer}$.
The limit $t\to 0$ gives
\beq
 \cz_{\ck}(1,\th^k) \to (1-1)\, (2t) \prod_{i=1}^3 |2 \sin 2\pi kv_i|
\label{z1klim}
\eeq
Also, taking into account sums over momenta/windings and using the Poisson
resummation formula \cite {gp}, we find
\beqa
\cz_{\ck}(1,1) & \to & (1-1)\, (16t)\, \frac{V_1V_2V_3}{(4\pi^2 \a^\prime)^3}
\nonumber \\[0.3ex]
\cz_{\ck}(1,R) & \to & (1-1)\, (16t)\, \frac{4\pi^2 \a^\prime V_3}{V_1V_2}
\label{mz1klim}
\eeqa
After the change of variables
$t=\frac1{4\ell}$  \cite{gp}, the $\ck$ amplitude schematically reduces to
$(1-1)\int_0^\infty d\ell \sum_k \tilde Q_k^2$. The $(1-1)$ shows that,
as expected, the NS-NS and R-R are equal and cancel in the full amplitude.
However, consistency of field equations \cite{pcai} requires that each
divergence vanishes separately but in general the orientifold plane charges
$\tilde Q_k$ are not zero as seen from the above limits for
$\cz_{\ck}(1,\th^k)$.

For $N$ even we also have to consider $\cz_{\ck}(R,\th^k)$. An interesting
example to analyze is $Z_4$. In this case $v_3=-\oh$ and we must include
a sum over windings in $Y_3$ in both
$\cz_{\ck}(1,\th^k)$ and $\cz_{\ck}(R,\th^k)$, $k=1,3$.
We obtain
\beq
\cz_{\ck}(1,\th^k) +  \cz_{\ck}(R,\th^k)
\to  (1-1)  \frac{16\pi^2 \a^\prime \, t}{V_3} [ (2\sin \frac{\pi k}2)^2
+ \tilde \chi(\th^2, \th^k) ] \quad ; \quad k=1,3
\label{zklim4}
\eeq
Here $\tilde \chi(\th^2, \th^k)=4$ is the number of simultaneous fixed
points of $\th^2$ and $\th$ (or $\th^3$). Thus, $\ck$ has a non-vanishing
tadpole proportional to $V_4/V_3$. In fact, whenever $\th^k$ reflects the
third coordinate we find that $\cz_{\ck}(1,\th^k) +  \cz_{\ck}(R,\th^k)$
leads to a non-zero tadpole proportional to
$V_4/V_3$. In $Z_8$, $Z_{12}^\prime$ this happens
for $k=\frac{N}4 {\rm \, mod \,} 2$,
and in $Z_8^\prime$, for $k=2,6$.
The $Z_2 \times Z_4$ and $Z_4 \times Z_4$ orientifolds
also have non-vanishing Klein-bottle tadpoles proportional
to $V_4/V_3$ that arise from the elements corresponding
to $v=(\frac14,\frac14,-\frac12)$ and $3v$.

To cancel the divergence proportional to $V_4V_1V_2V_3$ in $\ck_1(1)$
we must introduce 9-branes and to cancel the divergence proportional to
$V_4V_3/V_1V_2$ in $\ck_1(R)$ we must introduce $5_3$-branes. We then have
to compute M\"obius strip and cylinder amplitudes. These
amplitudes will also have divergences of the form
$(1-1)\int_0^\infty d\ell \sum_k \tilde Q_k Q_k$ and
$(1-1)\int_0^\infty d\ell \sum_k Q_k^2$. The various divergences from
all amplitudes can be organized according to volume dependence. For each
type of dependence we must find a sum of squares, each of which has to 
be zero.

The cylinder amplitudes are given by
\beq
\cc_{pq} = \frac{V_4}{8N} \sum_{k=0}^{N-1}  \int_0^\infty \, \frac{dt}t \,
(8\pi^2 \a^\prime t)^{-2} \, \cz_{pq}(\th^k)
\label{camp}
\eeq
where
\beq
\cz_{pq}(\th^k) = \Tr_{pq} \{(1+(-1)^F)  \th^k \,
{\rm e}^{-2\pi t L_0} \}
\label{zcdef}
\eeq
The trace is over open string states with boundary conditions according
to the D$p$ and D$q$-branes at the endpoints. $\cz_{pq}(\th^k)$ gives
rise to $\cc_{pq}(\th^k)$ with obvious definition from (\ref{camp}).

In $\cz_{99}$, boundary conditions are NN
in all directions. Hence,
\beq
\cz_{99}(\th^k) = \sum_{\a,\b=0,\oh}  \eta_{\a,\b} \
\frac{\vartheta[{\a \atop \b}]}{\eta^3} \ \prod_{i=1}^3
\  \frac{\vartheta[{\a \atop {\b + kv_i}}]}\eta  \,
\frac{-2\sin \pi k v_i \,  \eta}{\vartheta[{\oh \atop {\oh +kv_i}}]} \
(\Tr\g_{k,9})^2
\label{z991}
\eeq
This vanishes by supersymmetry. Indeed, using the first identity in
(\ref{abtru}) gives
\beq
\cz_{99}(\th^k) = (1-1) \, \frac{\vartheta[{0 \atop \oh}]}{\eta^3}
 \ \prod_{i=1}^3
\frac{-2\sin \pi k v_i \,\vartheta[{0 \atop {\oh + kv_i}}]}
{\vartheta[{\oh \atop {\oh +kv_i}}]} \
(\Tr \g_{k,9})^2
\label{z992}
\eeq
When $kv_i = {\rm integer}$ there is a well defined limit and we must include
a sum over quantized momenta in $Y_i$. In the limit $t \to 0$ we find
\beq
\cz_{99}(1)  \to  (1-1)\, t\, \frac{V_1V_2V_3}{(8\pi^2 \a^\prime)^3}
\, (\Tr \g_{0,9})^2
\label{c991lim}
\eeq
Also, if $kv_i \not= {\rm integer}$,
\beq
 \cz_{99}(\th^k) \to (1-1)\, t \, \prod_{i=1}^3 |2 \sin \pi kv_i| \,
(\Tr \g_{k,9})^2
\label{c99lim}
\eeq
However, if for instance $kv_3 = {\rm integer}$,
\beq
 \cz_{99}(\th^k) \to (1-1)\, t \, \frac{V_3}{8\pi^2 \a^\prime}
 \prod_{i=1}^2 |2 \sin \pi kv_i| \, (\Tr \g_{k,9})^2
\label{c99ilim}
\eeq
To extract the divergences in all $\cc$ amplitudes we have to make the
change of variables $t=\frac1{2\ell}$  \cite{gp}.

In the (55) sector there are DD boundary conditions in directions $Y_1,Y_2$
transverse to the 5-branes. Oscillator expansions with DD boundary conditions
have integer modes but include windings instead of momenta. Then, $\cz_{55}$
has a form similar to (\ref{z991}). More precisely, after using the first
identity in (\ref{abtru}),
\beq
\cz_{55}(\th^k) = (1-1) \, \frac{\vartheta[{0 \atop \oh}]}{\eta^3}
 \ \prod_{i=1}^3
\frac{-2\sin \pi k v_i \,\vartheta[{0 \atop {\oh + kv_i}}]}
{\vartheta[{\oh \atop {\oh +kv_i}}]} \
\sum_I \, (\Tr\g_{k,5,I})^2
\label{z55}
\eeq
where $I$ refers to the fixed points of $\th^k$.  This is valid if
$kv_i \not= {\rm integer}$. Otherwise we must include
a sum over windings in $Y_i , i=1,2$, or over quantized momenta in $Y_3$.
For example, for $k=0$ we find the $t \to 0$ limit
\beq
 \cz_{55}(1) \to (1-1)\, \frac{t}{16} \frac{8\pi^2 \a^\prime V_3}{V_1V_2}
\, (\Tr \g_{0,5})^2
\label{c551lim}
\eeq

In the (59) sector there are DN boundary conditions in coordinates $Y_1,Y_2$.
Hence, their oscillator expansions include half-integer modes. For fermions,
world-sheet supersymmetry requires  that in Neveu-Schwarz (Ramond), moddings
be opposite (same) to that of the corresponding bosons. Hence,
\beq
\cz_{59}(\th^k) = \sum_{\a,\b=0,\oh}  \eta_{\a,\b} \,
\frac{\vartheta[{\a \atop \b}]}{\eta^3}
\frac{-2\sin \pi k v_3 \vartheta[{\a \atop {\b + kv_3}}]}
{\vartheta[{\oh \atop {\oh +kv_3}}]} \,
\prod_{i=1}^2
\  \frac{\vartheta[{{\a+\oh} \atop {\b + kv_i}}]}
{\vartheta[{0 \atop {\oh +kv_i}}]} \,
\Tr\g_{k,9} \sum_I \, \Tr\g_{k,5,I}
\label{z591}
\eeq
Using the second identity in (\ref{abtru}) shows that, as expected,
$\cz_{59}$ vanishes and can be written as
\beq
\cz_{59}(\th^k) = (1-1) \, \frac{\vartheta[{0 \atop \oh}]}{\eta^3}
\frac{-2\sin \pi k v_3 \, \eta}{\vartheta[{\oh \atop {\oh +kv_3}}]} \,
\frac{\vartheta[{0 \atop {\oh +kv_3}}]}{\eta}
 \ \prod_{i=1}^2
\frac{\vartheta[{\oh \atop {\oh + kv_i}}]}
{\vartheta[{0 \atop {\oh +kv_i}}]}\,
\Tr\g_{k,9} \sum_I \,\Tr\g_{k,5,I}
\label{c592}
\eeq
Notice that for $k=0$, or for $kv_2={\rm integer}$ as in $Z_6^\prime$,
the expression multiplying $(1-1)$ always vanishes.
For $k=\frac{N}2$ there is a well defined limit and we must include a sum
over quantized momenta in $Y_3$.

The M\"obius strip amplitudes are given by
\beq
\cam_p = \frac{V_4}{8N} \sum_{k=0}^{N-1}  \int_0^\infty \, \frac{dt}t \,
(8\pi^2 \a^\prime t)^{-2} \, \cz_p(\th^k)
\label{mamp}
\eeq
where
\beq
\cz_p(\th^k) = \Tr_p \{(1+(-1)^F) \, \Om \, \th^k \,
{\rm e}^{-2\pi t L_0} \}
\label{zmdef}
\eeq
$\cz_p(\th^k)$ gives rise to $\cam_p(\th^k)$ with obvious definition
from (\ref{mamp}).
Here the trace is over open string states with boundary conditions
according to the D$p$-branes at both endpoints. The main difference
between $\cz_p$ and $\cz_{pp}$ is the insertion of $\Omega$ that
acts on the various bosonic and fermionic oscillators thereby
introducing extra phases in the expansions in $q$. More precisely,
$\Omega$ acts on oscillators as \cite{gp}
\beq
\a_r \to  \pm {\rm e}^{i\pi r}
\quad\quad ; \quad\quad
\psi_r \to  \pm {\rm e}^{i\pi r}
\label{apsi}
\eeq
The upper (lower) sign is for NN (DD) boundary conditions.
Furthermore, $\Omega$
acts as ${\rm e}^{-i\pi/2}$ on the NS vacuum. This ensures that
$\Om( \psi^\mu_{-\oh} \vac_{NS})  =  - \psi^\mu_{-\oh} \vac_{NS}$ as needed
for the orientifold projection on gauge vectors.

To derive the M\"obius trace we can use (\ref{apsi}) and the results for
(99) cylinders. After using $\vt$ identities we obtain
\beq
\cz_9(\th^k) = -(1-1) \,
\frac {\tilde\vt[{\oh \atop 0}]\, \tilde\vt[{0 \atop \oh }]}
{\tilde\eta^3 \tilde\vt[{0 \atop 0}]} \ \prod_{i=1}^3 \,
\frac{-2\sin \pi k v_i \,
\tilde\vt[{\oh \atop {kv_i}}]\, \tilde\vt[{0 \atop {\oh + kv_i}}]}
{\tilde\vt[{\oh \atop {\oh + kv_i}}] \, \tilde\vt[{0 \atop {kv_i}}]}
\, \Tr(\g_{\Omega_k,9}^{-1} \cdot \g_{\Omega_k,9}^T)
\label{zm9}
\eeq
where again the tilde means variable $\tilde q={\rm e}^{-4\pi t}$.
When $kv_i ={\rm integer}$ we must include a sum over quantized
momenta in $Y_i$.
Notice that $\tilde\vt[{\oh \atop {kv_i}}]$ vanishes identically when
$k=\frac{N}2$
and whenever $kv_i={\rm half}$-integer.
For $5_3$-branes we instead find
\beqa
\cz_5(\th^k) & = & (1-1) \,
\frac {\tilde\vt[{\oh \atop 0}]\, \tilde\vt[{0 \atop \oh }]}
{\tilde\eta^3 \tilde\vt[{0 \atop 0}]} \,
\frac{-2\sin \pi k v_3 \,
\tilde\vt[{\oh \atop {kv_3}}]\, \tilde\vt[{0 \atop {\oh + kv_3}}]}
{\tilde\vt[{\oh \atop {\oh + kv_3}}] \, \tilde\vt[{0 \atop {kv_3}}]}
\nonumber \\[0.3ex]
& {} & \prod_{i=1}^2 \,
\frac{2\cos \pi k v_i \,
\tilde\vt[{\oh \atop {\oh + kv_i}}]\, \tilde\vt[{0 \atop kv_i}]}
{\tilde\vt[{\oh \atop kv_i}] \, \tilde\vt[{0 \atop {\oh + kv_i}}]}
\, \sum_I \Tr(\g_{\Omega_k,5,I}^{-1} \cdot \g_{\Omega_k,5,I}^T)
\label{zm5}
\eeqa
For $k=0$ there is a vanishing contribution to tadpoles. For
$kv_i={\rm half}$-integer, $i=1,2$, we must include a sum over windings
in $Y_i$. For $kv_3={\rm integer}$ we must include a sum over momenta in
$Y_3$.
In particular, for $k=\frac{N}2$ we obtain the following $t \to 0$ limit
\beq
\cz_5(R) \to -(1-1)\, t \, \frac{8\pi^2 \a^\prime V_3}{V_1V_2}
\, \Tr \g_{0,5}
\eeq
To extract tadpoles in $\cam_p$ we have to make the change of variables
$t=\frac1{8\ell}$ \cite{gp}.

To arrive at the tadpole cancellation
conditions written in section~\ref{tadsec},
we must take the limit $t \to 0$ in the various
traces and next change variable to $\ell$ appropriately to find the large
$\ell$ behavior of the amplitudes. The final step is to collect all terms
with a given volume dependence.

To finish this appendix we wish to stress that in $Z_4, Z_8, Z_8^\prime$ and
$Z_{12}^\prime$ there are left-over tadpoles even after introducing 9-branes
and $5_3$-branes. Indeed, we have seen that in these cases the $\ck$ amplitude
has divergences proportional to $V_4/V_3$. This type of volume dependence
cannot arise from any of the $\cam$ or $\cc$ amplitudes because it would
require a sum over windings in $Y_3$ that is not possible for $5_3$-branes.

\bigskip\medskip
\noindent
{\large \bf Compendium of $\vt$ properties}
\medskip

The $\vt$ function of rational characteristics $\d$ and $\vphi$ is given by
\beq
\vt[{\d \atop \vphi}](t) = \sum_n q^{\oh(n+\d)^2} \,
{\rm e}^{2i\pi (n+\d) \vphi}
\label{vts}
\eeq
Here the variable $q$ is $q={\rm e}^{-2\pi t}$.
The $\vt$ function also has the product form
\beq
\frac{\vt[{\d \atop \vphi}]}{\eta} = {\rm e}^{2i\pi \d \vphi}
\, q^{\oh \d^2 - \frac1{24}} \,
\prod_{n=1}^\infty (1 + q^{n+\d -\oh} {\rm e}^{2i\pi \vphi} ) \,
(1 + q^{n-\d -\oh} {\rm e}^{-2i\pi \vphi} )
\label{vtp}
\eeq
where the Dedekind $\eta$ function is
\beq
\eta =  q^{\frac1{24}} \,
\prod_{n=1}^\infty (1 - q^n)
\label{deta}
\eeq
Notice that
\beq
\lim_{\vphi \to 0} \,
\frac{-2 \sin \pi \vphi }{\vt[{\oh \atop {\oh+\vphi}}]} =
\frac1{\eta^3}
\label{limu}
\eeq
The $\vt$ and $\eta$ functions have the modular transformation properties
\beqa
\vt[{\d \atop \vphi}](t) & = &{\rm e}^{2i\pi \d \vphi} \, t^{-\oh} \,
\vt[{-\vphi \atop \d}](1/t)
\nonumber \\[0.2ex]
\eta(t) & = &  t^{-\oh} \, \eta(1/t)
\label{modt}
\eeqa
The $\vt$'s satisfy several Riemann identities \cite{mum}. In particular,
\beqa
& {} & \sum_{\a,\b} \eta_{\a,\b} \  \vt[{\a \atop \b}] \prod_{i=1}^3
\vt[{\a \atop {\b+u_i}}] = 0
\nonumber \\[0.2ex]
& {} & \sum_{\a,\b} \eta_{\a,\b} \ \vt[{\a \atop \b}]
\vt[{\a \atop {\b+u_3}}]
\prod_{i=1}^2 \vt[{{\a+\oh} \atop {\b+u_i}}] = 0
\label{abtru}
\eeqa
provided that $u_1+u_2+u_3=0$.


\newpage

\begin{thebibliography}{99}
%
\bibitem{sagnotti}
A.~Sagnotti, in Cargese 87,
{\it Strings  on Orbifolds},
ed. G. Mack et al. (Pergamon Press, 1988) p. 521.
%
\bibitem{hor}
P.~Horava, \NPB{327} {89} {461}; \PLB{231} {89} {251};\\
J.~Dai, R.~Leigh and J.~Polchinski, Mod.Phys.Lett. A4 (1989) 2073;\\
R.~Leigh, Mod.Phys.Lett. A4 (1989) 2767.
%
\bibitem{bs} G.~Pradisi and A.~Sagnotti, \PLB{216} {89} {59};\\
M.~Bianchi and A.~Sagnotti, \PLB{247} {90} {517}.
%
\bibitem{gp}
E.~Gimon and J.~Polchinski,
Phys.Rev. D54 (1996) 1667, hep-th/9601038.
%
\bibitem{dp1}
A.~Dabholkar and J.~Park, \NPB{472} {96} {207}, hep-th/9602030.
%
\bibitem{gj}
E.~Gimon and C.~Johnson,
\NPB{477}{96}{715}, hep-th/9604129.
%
\bibitem{dp2}
A.~Dabholkar and J.~Park, \NPB{477}{96}{701}, hep-th/9604178.
%
\bibitem{dp3}
A.~Dabholkar and J.~Park, \PLB{394}{97}{302}, hep-th/9607041.
%
\bibitem{bz}
J.~Blum and A.~Zaffaroni, \PLB{387} {96} {71}, hep-th/9607019.
%
\bibitem{blum}
J.~Blum, \NPB{486}{97}{34}, hep-th/9608053.
%
\bibitem{kak4}
Z. Kakushadze, G. Shiu and S.-H. Tye, hep-th/9803141 .
%
\bibitem{poldb}
J. Polchinski, \PRL{75} {95}{4724}
%
\bibitem{bl}
M.~Berkooz and R.~G.~Leigh, \NPB{483} {97} {187}, hep-th/9605049.
%
\bibitem{ang}
C.~Angelantonj, M.~Bianchi, G.~Pradisi, A.~Sagnotti and Ya.S.~Stanev,
\PLB{385} {96} {96}, hep-th/9606169.
%
\bibitem{kak1}
Z.~Kakushadze, \NPB {512} {98} 221, hep-th/9704059.
%
\bibitem{kak2}
Z.~Kakushadze and G.~Shiu, \PRD{56} {97} {3686}, hep-th/9705163.
%
\bibitem{kak3}
Z.~Kakushadze and G.~Shiu, hep-th/9706051.
%
\bibitem{fin}
L.E.~Ib\'a\~nez, hep-th/9802103.
%
\bibitem{zwart}
G.~Zwart, hep-th/9708040.
%
\bibitem{odri}
D.~O'Driscoll, hep-th/9801114.
%
\bibitem{polten}
J.~Polchinski, \PRD {55} {97} {6423}, hep-th/9606165.
%
\bibitem{vafadt}
C.~Vafa, \NPB{273} {86} {592}.
%
\bibitem{fiq}
A.~Font, L.E.~Ib\'a\~nez, F.~Quevedo, \PLB{217}{89}{272}.
%
\bibitem{gopm}
R. Gopakumar and S. Mukhi, \NPB{479} {96}{260} , hep-th/9607041.
%
\bibitem{afiuv}
G.~Aldazabal, A.~Font, L.E.~Ib\'a\~nez, A.M.~Uranga and G.~Violero,
hep-th/9706158.
%
\bibitem{dhvw}
L.~Dixon, J.A.~Harvey, C.~Vafa and E.~Witten, \NPB{274} {86} {285}.
%
\bibitem{pcai}
Y.~Cai and J.~Polchinski, \NPB{296} {88} {91}.
%
\bibitem{orbi}
L.E.~Ib\'a\~nez, J.~Mas, H.P.~Nilles and F.~Quevedo,
\NPB{301}{88}{157};\\
A.~Font, L.E.~Ib\'a\~nez, F.~Quevedo and A.~Sierra,
\NPB{331}{90}{421}.
%
\bibitem{sen05150}
A.~Sen, \NPB{475} {96} {562}, hep-th/9605150.
%
\bibitem{sen11186}
A.~Sen, \NPB{489} {97} {139}, hep-th/9611186.
%
\bibitem{anton2}
I. Antoniadis, H. Partouche and T.R. Taylor, Nucl. Phys. Proc.
Suppl. 61A (1998) 58,
hep-th/9706211.
%
\bibitem{intri}
K.~Intrilligator, \NPB {496} {97} {177}, hep-th/9702038.
%
\bibitem{blumintri}
J.~Blum and K.~Intrilligator, \NPB{506} {97} 199, hep-th/9705044.
%
\bibitem{ant}
I.~Antoniadis, C.~Bachas, C.~Fabre, H.~Partouche and T.R.~Taylor,
\NPB{489} {97} {160}, hep-th/9608012.
%
\bibitem{gs}
M. Green and J. Schwarz, \PLB{149} {84}{117}.
%
\bibitem{sagcan}
A.~Sagnotti, \PLB{294} {92} {196}, hep-th/9210127.
%
\bibitem{mum}
D.~Mumford, {\it The Tata Lectures on Theta I}, Birkh\"auser, 1983.
%







\end{thebibliography}
\end{document}